\documentclass[a4paper]{article}
\usepackage{enumitem}

\usepackage{authblk, fancyhdr}
\usepackage[utf8]{inputenc}
\usepackage[T1]{fontenc}
\usepackage{graphicx}
\usepackage{hyperref}
\usepackage{textgreek}
\usepackage{amsmath}
\usepackage{amssymb}
\usepackage{lipsum}
\usepackage{float}
\usepackage[left=25mm, right=25mm, top=20mm, bottom=20mm, footskip=20pt, headheight=20pt, headsep=0pt]{geometry}
\usepackage{subcaption}
\usepackage{booktabs}
\usepackage{tabularx}
\usepackage{makecell}
\usepackage{threeparttable}
\usepackage{amssymb}   
\usepackage[table]{xcolor}

\definecolor{lightblue}{RGB}{235,244,252}
\fancypagestyle{firstpage}{%
  \fancyhf{}
  \fancyhead[R]{}
  \fancyfoot[C]{\thepage}
}

\providecommand{\keywords}[1]
{
  \vspace{5pt}	
  \textbf{Keywords	} #1
}

\providecommand{\abstractnew}[1]
{
  \textbf{Abstract	} #1
}
\begin{document}
\pagenumbering{arabic}
\title{Temporal Boundaries of Newell's Car-Following Model: Insights from Lane-Free Traffic}

\author[1]{Suhaib Nazir\thanks{Email of the corresponding author: \url{suhaib@smail.iitm.ac.in}}}
\author[2,1]{Hillel Bar-Gera}
\author[1]{Bhargava Rama Chilukuri}

\affil[1]{Indian Institute of Technology Madras}
\affil[2]{Ben-Gurion University of the Negev}

\date{}
\maketitle\thispagestyle{firstpage}

\abstractnew
Newell's simplified car-following model offers behavioural interpretability with minimal parameters, yet the temporal limits within which it holds during actual car-following interactions, particularly in lane-free traffic, remain poorly understood. This has direct consequences for Newell parameter estimation. This study evaluates the model using high-resolution UAV trajectories from a lane-free highway corridor in Chennai, applying two independent estimation approaches: linear regression of the speed–spacing relationship (aggregate and pair-specific), and a shifting-optimization method that recovers parameters by minimizing spacing variance. A boundary-corrected variant of the shifting method is later introduced to isolate transitional regimes at interaction endpoints,  while preserving the original Newell formulation.
Pair-specific regression substantially outperforms the aggregate specification ($R^2$ = 0.84 vs. 0.51), underscoring pronounced driver heterogeneity. Parameters recovered via trajectory shifting are statistically equivalent to regression estimates, offering independent support for the trajectory-translation principle underlying Newell's model. Boundary correction markedly improves fit, raising mean $\bar{R^2}$ from 0.66 to 0.95 and being preferred for 81\% of pairs under BIC. The results point to interaction boundary effects, not failures in the core behavioural assumptions, as the principal source of estimation error in this setting. Correcting for them improves parameter recovery without compromising model parsimony. Thus, boundary correction is a necessary step in trajectory-based calibration of Newell-type models, with implications for traffic-state estimation and microscopic simulation.


\section{Introduction}
The mathematical representation of microscopic vehicle interactions serves as the fundamental mechanism by which individual driving behaviours are translated into systemic traffic phenomena \cite{KASHIFI2026103399, zhou2025twenty}. At the core of these interactions is the car-following (CF) process, which characterizes the longitudinal interaction between a vehicle and its immediate predecessor, a process governed by complex psychological, kinematic, and physical constraints\cite{zhang2024car, chen2024metafollower}. Over the past seven decades, research has progressed from simple stimulus-response formulations, such as Pipes \cite{pipes1953operational} and early General Motors (GM) models \cite{herman1959single}, to sophisticated psychophysical and desired-behaviour frameworks, such as the Intelligent Driver Model (IDM) \cite{treiber2000congested} and now the Intelligent Agent Model (IAM) \cite{treiber2022intelligent}. A fundamental conceptual distinction, however, runs through this body of work. Certain formulations, notably the IDM and Wiedemann's psychophysical model \cite{wiedemann1974simulation}, are designed to reproduce the full spectrum of longitudinal driving behaviour, spanning free-flow conditions, the approach phase, constrained following, and emergency braking. Others are explicitly scoped to the core car-following regime: the condition in which a follower is continuously constrained by its immediate leader and maintains an equilibrium spacing. These two model classes rest on different theoretical premises, operate over different domains of applicability, and consequently require different standards of empirical validation.

Among the existing modelling paradigms, Newell’s simplified car-following model \cite{NEWELL2002195} belongs to the latter category, and remains one of the most significant contributions to the field due to its parsimony and behavioural simplicity \cite{chen2012behavioral, li2020trajectory}. Newell’s formulation operates on a lower-order kinematic principle, assuming that in congested conditions, a following vehicle replicates its leader’s trajectory, subject only to a fixed translation in time and space. This regime-specific scope carries a direct methodological implication: the model's validity can be meaningfully assessed only against observations drawn exclusively from the core CF interval. Prior validation studies in homogeneous, lane-disciplined settings have addressed this requirement inconsistently; some impose headway or relative-speed thresholds to screen candidate following intervals \cite{ahn2004verification, duret2008estimating}, while others apply no explicit filtering at all \cite{chiabaut2010heterogeneous, zheng2012evaluation}. No consensus procedure for isolating the true CF interval has emerged, and the sensitivity of validation outcomes to this definitional choice has rarely been examined \cite{kesting2008calibrating}. This gap becomes considerably more acute once lane structure is absent. In lane-free, heterogeneous traffic, the following process itself grows substantially more complex, and even the preliminary task of identifying genuine CF episodes becomes challenging. This constitutes a methodological gap that directly undermines the validity of any subsequent model assessment.

Owing to its simplicity and computational efficiency, Newell's model has been widely applied in large-scale traffic simulation, traffic state estimation, and trajectory reconstruction \cite{li2020trajectory, zhang2024car}. However, like most classical CF models, its formulation rests on an implicit assumption of lane-based following, wherein each follower is constrained by a single leader and vehicle interactions occur primarily along the longitudinal direction \cite{KASHYAPNR2024129866, zhou2025twenty}. This assumption is increasingly challenged by the reality of heterogeneous, lane-free traffic where vehicles of varying dimensions and dynamic capabilities interact laterally as well as longitudinally.
Disordered or mixed traffic is characterized by two interrelated features: a highly heterogeneous vehicle composition and the absence of organized lane structures. Under such conditions, vehicle interactions are no longer governed solely by longitudinal dynamics but are strongly influenced by complex lateral movements, irregular vehicle arrangements, and continuously changing neighbouring vehicles, making the direct application of conventional CF models considerably more challenging \cite{KASHYAPNR2024129866, CFmodelsforHD&AV}.

This paper makes contributions at two levels: methodological and empirical. Methodologically, we address this gap by developing a coherent, reproducible procedure for isolating core car-following episodes in lane-free traffic, formalized around explicit criteria for leader--follower pair selection, following-interval boundaries, and minimum data requirements. Building on this foundation, we develop a quantitative, multi-method framework for assessing the validity of Newell's car-following principles, in which convergent evidence from complementary analytical approaches replaces reliance on any single criterion. Both components are documented in sufficient detail to support replication in comparable settings.
The empirical contributions stem from applying these methods to a high-resolution microscopic trajectory dataset collected under lane-free traffic conditions. The analysis yields a systematic evaluation of Newell's model in this context, establishing where the classical formulation holds and where it fails. We identify and characterize the behavioural regimes responsible for observed deviations from the original model, and use this characterization to motivate a simple extension that accommodates these departures while preserving the interpretability and core structure of the Newell formulation.

\section{Literature Review}
The evolution of traffic flow modelling reflects a continuous effort to balance behavioural realism with computational efficiency. Early analytical models, such as the Pipes \cite{pipes1953operational} and GM models \cite{herman1959single}, were particularly valued because their parameters had clear physical meanings, enabling direct understanding of driver response, desired spacing, and stability mechanisms. As modelling approaches advanced, richer behavioural representations such as psycho-physical models \cite{wiedemann1974simulation} and the IDM \cite{treiber2000congested} were introduced to capture heterogeneity and complex interactions, often improving predictive performance but reducing interpretability. More recently, with the rise of data-driven and machine learning methods, the literature has increasingly emphasized explainability techniques, hybrid physics-informed frameworks, and feature attribution methods to recover behavioural meaning while retaining accuracy. This section organizes the existing body of knowledge thematically, focusing on the historical progression of car-following theory (section \ref{sec:cf models}), the specialized paradigm of Newell’s model (section \ref{sec:nw paradigm}), pair identification methodologies (section \ref{sec:lf identification}) that are important for the analysis in this paper, and finally in (section \ref{sec:research need}) discusses the need for this study.

\subsection{Foundations and Evolution of Car-Following Models} \label{sec:cf models}
The systematic study of car-following behaviour dates to the mid-twentieth century. Early models by Reuschel \cite{reuschel1950fahrzeugbewegungen} and Pipes \cite{pipes1953operational} laid the groundwork by proposing that a following vehicle maintains a safe distance proportional to its speed. This stimulus–response logic was formalized by Chandler et al. \cite{chandler1958traffic} who framed a driver's acceleration as a reaction to the relative speed of the vehicle ahead, occurring after a finite time lag. This was later generalized by the Gazis–Herman–Rothery (GHR) model \cite{gazis1961nonlinear}, in which acceleration was made proportional to relative speed and inversely proportional to spacing. Much of the subsequent literature focused on how the reaction lag influences traffic stability, particularly the amplification of disturbances as they propagate upstream. Despite their influence, these early formulations struggled to reproduce congested-regime phenomena such as stop-and-go waves \cite{treiber2000congested}, motivating a shift toward models grounded in explicit safety or behavioural criteria rather than stimulus–response calibration alone.

The Gipps model \cite{gipps1981behavioural} addressed this by deriving the follower's speed directly from a collision-avoidance criterion, i.e., the speed that would still allow a safe stop if the leader braked at maximum deceleration. Since every parameter corresponds to a measurable physical or psychological driver characteristic, its appeal lies in interpretability. However, this same specificity makes the model heavily parameterized, and calibrating it against real-world trajectory data was challenging \cite{ossen2005car}.

A parallel line of work sought smoother, more continuous alternatives to these threshold-based safety rules. The Optimal Velocity Model (OVM) \cite{bando1995dynamical} recast CF as a relaxation toward a speed–spacing equilibrium and was the first model of this family shown analytically to reproduce the spontaneous formation of stop-and-go waves. Because the assumption of instantaneous velocity adjustment in OVM could produce unrealistically large accelerations, Helbing and Tilch \cite{helbing1998generalized} introduced the Generalized Force Model (GFM), adding an explicit deceleration term for negative speed differences and calibrating it against floating-car data. These optimal-velocity-type models established the desired-behaviour paradigm, governing acceleration through a target state rather than a safety envelope, that later models would build on. Later, desired-behaviour models like the IDM gained prominence for their ability to reproduce smooth transitions between free-flow and congested regimes using intuitive parameters such as desired time headway and maximum acceleration \cite{zhang2024car}. It has since become a standard simulation benchmark. The model is, however, deterministic and produces trajectories that are too smooth to represent the aggressive or inconsistent behaviour observed in real traffic \cite{Zhouzheng}. The IAM \cite{treiber2022intelligent} extends the IDM further, from a one-dimensional longitudinal axis to a continuous two-dimensional coordinate field, by coupling its CF mechanics with the force-based philosophy of Social Force Models \cite{helbing1995social}. Lateral displacements emerge from spatially resolved force fields rather than discrete lane-assignment rules, allowing the IAM to represent both lane-based and lane-free traffic within a single formulation. However, the generality comes at a high computational cost.

Across the discussed models (e.g., IDM, GHR, IAM), driver behaviour is expressed as an acceleration response. Calibrating these models against field data, therefore, requires acceleration estimates obtained by differentiating measured positions or speeds. This amplifies any measurement noise present in the underlying trajectory data, and this amplification compounds at each successive order of differentiation \cite{punzo2011assessment}. Left uncorrected, this error propagation can distort calibrated parameter values. Beyond this, the parameters themselves pose a further difficulty for calibration. These models take a mechanistic or behavioural modelling stance, attempting to represent the internal control logic of the driver, which makes them less intuitive to work with. While parameters such as desired speed have a clear physical interpretation, others, such as the acceleration exponent and comfortable deceleration, are latent behavioral constructs with no direct empirical counterpart. Their estimation typically requires intensive numerical optimization, which can yield multiple solution.

\subsection{Newell’s Paradigm} \label{sec:nw paradigm}
Newell’s simplified car-following model employs a fundamentally different behavioural logic from earlier formulations and requires substantially less complexity than preceding models \cite{NEWELL2002195, li2020trajectory}. The model is also less susceptible to measurement noise, since it is formulated in terms of positions and velocities rather than accelerations. Newell's parameters are empirically transparent, meaning they are physical measurements that can be estimated directly from observed wave propagation. 

Newell's core hypothesis is that a driver selects a preferred spacing for each velocity such that the follower's trajectory is a translation of the leader's trajectory, offset by fixed constants in time and space, as illustrated in Fig. \ref{fig:newell}.  It represents longitudinal vehicle interactions through a simple mathematical framework.
\begin{equation}
   x_{n}(t) = x_{n-1}(t - \tau_{n}) - \delta_{n}
\end{equation}

\noindent where $x_n(t)$ represents the position of the follower vehicle and $x_{n-1}(t)$ that of its leader at time $t$; $\tau$ denotes the time offset or time lag, and $\delta$ denotes the space offset or jam spacing driver is willing to accept (depends on vehicle length and the driver's comfort buffer).

In the original microscopic formulation, Newell's parameters are driver-specific but vary across different drivers \cite{NEWELL2002195, chen2012behavioral}.  Newell conjectured that these parameters vary between individuals as if they were independently sampled from a joint probability distribution \cite{wang2008effect}, reflecting the fact that some drivers prefer to follow closely while others maintain a larger cushion. However, when Newell translated these microscopic variables into macroscopic variables, he replaced the driver-specific parameters with their statistical means, essentially homogenizing them to represent an average driver \cite{jabari2014probabilistic}. It should be noted that homogeneity is an idealized property as it implies identical drivers.

Several studies have empirically verified Newell's linearity and fixed-parameter assumptions using microscopic vehicle trajectory data. Ahn et al. \cite{ahn2004verification} constructed piecewise linear trajectories for vehicles discharging from signalized intersections and traced speed waves at 0, 6.5, 13, and 19.5 km/h. Maximum likelihood estimation and likelihood ratio tests confirmed that the temporal and spatial translations between consecutive trajectories followed a common bivariate normal distribution independent of velocity. They also found that drivers do not adopt the leader's velocity after a fixed reaction time but rather when spacing reaches the value they associate with that velocity, a subtler relationship than simple temporal lag. Wang and Coifman \cite{wang2008effect} tested linearity at the individual vehicle level using Pearson correlation coefficients between spacing and speed. Most vehicles showed strong linear trends, and the relationship strengthened once transient data points from lane-changing maneuvers were removed. Ma and Ahn \cite{ma2008comparisons} covered a wider range of congested speeds (0–88 km/h) by averaging observations at 5~s intervals to reduce variance before applying generalized least squares regression across multiple freeway lanes. In the absence of lane changes, the linear speed-spacing relationship was statistically significant in every lane, suggesting a single linear model can broadly characterize car-following behavior. Duret et al. \cite{duret2008estimating} investigate inter-driver variability during congestion through the speed--space relationship, focusing on two parameters: maximum wave speed and minimum (jam) spacing. Comparing a data-driven estimation procedure with a simulation-based optimization of Newell's CF model, the study establishes that drivers do not conform to a single homogeneous rule and that explicitly incorporating driver-specific parameters substantially reduces trajectory prediction error. However, the study has a high data exclusion rate due to inconsistent values and the jam spacing estimates are unrealistic. On the fixed-parameter side, Chiabaut et al. \cite{chiabaut2010heterogeneous} tried minimizing the standard deviation of spatial shifts along NGSIM trajectories to recover the optimal wave speed. Splitting trajectories into equal time periods, they confirmed that the parameters $\tau$ and $\delta$ vary across drivers but remain stable within each individual over time.
Several studies have relaxed Newell's assumption of constant within-driver parameters to account for more dynamic and stochastic traffic behaviour. Ahn et al. \cite{ahn2013method} allowed both the spacing and time parameters to vary over time, enabling the model to reproduce hysteresis and non-steady-state conditions that a fixed-parameter formulation cannot capture. Laval and Leclercq \cite{laval2010mechanism} introduced a behavioural deviation term and showed that the interplay between timid and aggressive drivers is sufficient to trigger spontaneous traffic jams. Chen et al. \cite{chen2012behavioral, chen2012microscopic} pursued this line further with an asymmetric model that distinguishes driver reactions across the acceleration and deceleration phases of traffic oscillations. More recent contributions have moved toward explicitly stochastic formulations. For example, Meng et al. \cite{meng2021modification} added stochastic components to the steady-state Newell equation to account for vehicle dynamics and driving-situation-dependent variability, applying the resulting model to emission estimation. Although these extensions relax the assumption that Newell's within-driver parameters are fixed and deterministic, the original formulation remains the necessary baseline from which any further modifications should be empirically justified.

The major advantage of Newell’s model is its parsimony. It characterizes each LF pair with only two parameters compared to the five or more parameters required by models such as the IDM or Wiedemann. This simplicity facilitates analytical tractability and renders the model well-suited to large-scale network applications \cite{taylor2015method}. Empirical evidence also indicates that, in the absence of heavy lane-changing and complex geometry, the model remarkably and accurately predicts general car-following behavior across a massive range of congested speeds \cite{wang2008effect, ma2008comparisons}. However, the deterministic nature of the classical Newell model introduces several limitations. Because this deterministic framework forces disturbances to propagate at a constant wave speed without amplifying or decaying, the model fundamentally fails to capture traffic instability phenomena, such as the spontaneous formation of stop-and-go waves or the naturally occurring concave growth pattern of traffic oscillations \cite{chen2012behavioral, ZHENG2023104276}. The extent to which this deterministic structure captures the inherent intra-driver heterogeneity observed in real traffic remains unclear, especially in disordered traffic conditions. Furthermore, empirical evidence suggests that reaction time may not remain strictly constant over the course of a trip \cite{ZHENG2023104276,tian2021car}, and that driver behaviour can vary across different traffic regimes. Extensions of the model have explored relaxing this assumption, for instance by introducing stochastic wave travel times or imposing upper bounds on acceleration, to better reproduce oscillation growth \cite{ZHENG2023104276}. Despite these shortcomings, the model's limitations are largely outweighed by its profound parsimony and analytical utility. Compared to highly complex stimulus-response or psychophysical models, Newell’s framework relies on very few, easily interpretable parameters \cite{ma2008comparisons}. Moreover, its mathematical elegance enables easy extension to capture driver heterogeneity and hysteresis without sacrificing the model's core computational efficiency. The summary of various studies evaluating Newell's CF model is given in Table \ref{tab:lit_review}.

\begin{table*}[!ht]
\centering
\caption{Comparison of studies evaluating Newell's simplified car-following model}
\label{tab:lit_review}
\renewcommand{\arraystretch}{1.2}
\begin{threeparttable}
\resizebox{\textwidth}{!}{
\begin{tabular}{p{2cm} p{2.5cm} p{3.5cm} p{4cm} p{2.8cm} p{4cm} p{4cm}}
\hline
\rowcolor{blue!20}
\textbf{Study} &
\textbf{Traffic/ Data type} &
\textbf{Objective} &
\textbf{Methodology} &
\textbf{Performance measure} &
\textbf{Key findings} &
\textbf{Remarks} \\
\hline

\rowcolor{blue!5}
Newell \cite{NEWELL2002195} &
Theoretical &
Propose simplified CF model &
Mathematical derivation of TT hypothesis &
Analytical formulation &
Constant time--space translation; linear S--S relationship &
Did not validate empirically \\

Ahn et al. \cite{ahn2004verification} &
Urban signals (queue discharge) &
Verify simplified CF theory &
Estimated shock-wave speeds; compared across speed levels &
Statistical comparison of wave speeds &
Constant wave speed and linear congested branch &
Queue discharge at signals, not in continuous freeway traffic; vehicles sampled in acceleration states only \\

\rowcolor{blue!5}
Wang \& Coifman \cite{wang2008effect} &
Freeway with lane LC (NGSIM) &
Examine microscopic validity \& LC effects &
S--S correlation for individual vehicles before/after LC &
Pearson correlation coefficient &
Linear S--S relationship; LC disrupts equilibrium &
Did not verify TT \\

Duret et al. \cite{duret2008estimating} &
Congested freeway (NGSIM I-80) &
Estimate S--S relationships; assess Newell's ability to reproduce follower trajectories &
Individual pair S--S linear regression; LV problem to estimate individual parameters &
$R^2$, RMSE &
S--S relationships \& TT vary across vehicles; parameter distributions reveal driver heterogeneity &
No LF pair estimation; unrealistic jam spacing estimates \\

\rowcolor{blue!5}
Ma and Ahn \cite{ma2008comparisons} &
Congested freeways (NGSIM I-80 and US-101) &
Extend verification over wider speed range and LC conditions &
S--S regression ; comparison across lanes and during LC &
$R^2$, statistical tests &
Linear S--S relationship across lanes; quantified anticipation and relaxation periods &
Relied on S--S relationship only \\

Chiabaut et al. \cite{chiabaut2010heterogeneous} &
Congested freeway (NGSIM I-80) &
Estimate Newell parameters \& connect microscopic behaviour to macroscopic traffic &
Direct parameter estimation for stable LF pairs &
Parameter distributions; stationarity analysis &
Heterogeneous Newell parameters; linked microscopic parameters with macroscopic wave properties &
Parameter estimation under stable lane-based congestion \\

\rowcolor{blue!5}
Zheng et al. \cite{zheng2012evaluation} &
Urban freeway (NGSIM) &
Compare major CF models &
GA calibration \& validated several CF models including Newell &
Relative spacing error &
Newell performed competitively, but calibration and validation errors increased under different conditions &
Treated Newell as one of several simulation models; assumptions not validated \\

\rowcolor{blue!10}
This study &
Lane-free; heterogeneous; congested; Indian Highway &
Evaluate the validity of Newell's model and identify systematic deviations &
Verify S--S relationship; verify TT; estimate deviations &
$R^2$, $\bar{R}^2$, RMSE, parameter distributions &
Assumptions are valid for core CF regimes &
Comprehensive \& lane-free validation of Newell's model \\

\hline
\end{tabular}}
\begin{tablenotes}[flushleft]
\footnotesize
\item \textit{Note:} LC = lane change; S--S = speed--spacing; TT = trajectory translation; CF = car following
\end{tablenotes}
\end{threeparttable}
\end{table*}

\subsection{Leader--Follower Pair Identification} \label{sec:lf identification}
Identifying leader-follower (LF) pairs is a necessary precursor to any microscopic car-following analysis. The extraction of these pairs from trajectory data requires navigating a complex landscape of spatial and temporal thresholds that define when a vehicle is truly influenced by a predecessor. In lane-free traffic, this task is considerably more demanding than in homogeneous lane-based settings, where lane assignment implicitly resolves the pairing problem \cite{anil2022calibrating, anand2019calibration, madhu2020following, raju2021modeling}. Researchers have accordingly developed multi-criteria frameworks that jointly assess longitudinal clearance, lateral position, time headway, and interaction duration.
Longitudinal clearance delimits the outer boundary of the leader's influence zone. Thresholds vary with road type and traffic density, ranging from 30~m in dense urban conditions \cite{anand2019calibration} to 120~m on expressways \cite{zhu2018modeling}. Beyond these bounds, a trailing vehicle is generally regarded as operating in free-driving mode and therefore outside the scope of car-following behavior \cite{anil2022calibrating}.

Lateral criteria are particularly consequential in lane-free settings, where vehicles are unconstrained in their lateral positioning. A common approach imposes an explicit maximum lateral separation of 2.5 m \cite{zhu2018modeling} to ensure that two vehicles are traversing the same path. Some studies compute the lateral extents of each vehicle from trajectory data and require that the front lateral edge of the follower overlaps with the rear lateral edge of the leader, with a small safety buffer \cite{papathanasopoulou2018flexible,anil2022calibrating}. Recent work, however, has questioned whether strict physical overlap is a necessary condition. For example, Kulkarni et al. \cite{kulkarni2025leader} argue that vehicles separated by a small lateral gap may still exert a meaningful influence on one another.

Time headway provides a dynamic complement by incorporating vehicle speed into the identification process. A two-second threshold, for instance, is obtained by dividing a 30~m influence zone by an assumed average stream speed of 15~m/s \cite{anand2019calibration}. It should be noted, however, that the choice of cutoff is consequential: shifting from 2.0 s to 2.5 s can reclassify a substantial number of trajectory points from non-interacting to interacting status \cite{kulkarni2025leader}.

Beyond geometric and headway criteria, most identification frameworks impose a minimum continuous following duration, ranging from 5~s \cite{anil2022calibrating} to 30~s \cite{higgs2013two}  to eliminate spurious pairs. Additionally, some studies \cite{anil2022calibrating, raju2021modeling} verify the absence of an intervening vehicle, since a third vehicle breaking into the gap effectively dissolves the direct LF relationship. Table~\ref{tab:lf_criteria} summarizes these criteria across the studies reviewed.

\begin{table}[H]
\centering
\caption{Summary of LF pair identification criteria across selected studies}
\label{tab:lf_criteria}
\footnotesize
\begin{threeparttable}
\rowcolors{2}{blue!5}{white}
\renewcommand{\arraystretch}{1.3}
\begin{tabularx}{\linewidth}{ p{0.20\linewidth} p{0.16\linewidth} p{0.10\linewidth} X p{0.10\linewidth}
p{0.10\linewidth}}
\toprule
\rowcolor{blue!15}
\textbf{Study}
& \textbf{\makecell[l]{Traffic /\\road type}}
& \textbf{\makecell{Long.\\clearance}}
& \textbf{Lateral criterion}
& \textbf{\makecell{Time\\headway}}
& \textbf{\makecell{Following\\duration}} \\
\midrule
Zhu et al.~\cite{zhu2018modeling}
  & Expressway
  & <120~m
  & $\leq$2.5~m (lateral separation)
  & ---
  & ---\\[3pt]
Fernandez et al.~\cite{fernandez2011secondary}
  & Urban arterial
  & <100~m
  & ---
  & ---
  & --- \\[3pt]
LeBlanc et al.~\cite{leblanc2013longitudinal}
  & Lane-based
  & ---
  & ---
  & ---
  & >15~s \\[3pt]
Anand et al.~\cite{anand2019calibration}
  & Dense \& lane-free
  & <30~m
  & <3~m (lateral displacement)
  & <2~s
  & >5~s \\[3pt]
Anil et al.~\cite{anil2022calibrating}
  & Lane-free
  & <30~m
  & Overlap>0; strict (>50\% width), partial (<50\%)
  & ---
  & >5~s \\[3pt]
Papathanasopoulou \makecell{\& Antoniou~\cite{papathanasopoulou2018flexible}}
  & Lane-free
  & <200~m
  & Overlap + 0.2~m buffer
  & ---
  & --- \\[3pt]
Kulkarni et al.~\cite{kulkarni2025leader}
  & Lane-free
  & ---
  & Lateral clear gap (no overlap required)
  & 2.0--2.5~s\tnote{*}
  & --- \\[3pt]
Madhu et al.~\cite{madhu2020following}
  & Lane-free
  & <30~m
  & Lateral overlap$>0$
  & ---
  & --- \\[3pt]
Higgs \& Abbas~\cite{higgs2013two}
  & ---
  & 61--120~m
  & ---
  & ---
  & >30~s \\[3pt]
Raju et al.~\cite{raju2021modeling}
  & Lane-free
  & ---
  & <1.5~m (lateral clearance)
  & ---
  & ---\\
\bottomrule
\end{tabularx}
\begin{tablenotes}
\footnotesize
\item[*] Range reflects a sensitivity analysis (2.0~s vs.\ 2.5~s cutoff), not a single adopted threshold.
\end{tablenotes}
\end{threeparttable}
\end{table}

\subsection{Synthesis and Research Need} \label{sec:research need}
Newell's CF model occupies a distinctive position in traffic-flow theory, defined by its parsimony and behavioural interpretability. Empirical studies in lane-disciplined environments have confirmed that the trajectory-translation hypothesis reproduces congested following behaviour with considerable fidelity \cite{ahn2004verification,duret2008estimating}. Successive extensions, incorporating stochastic wave travel times \cite{meng2021modification, tian2021car}, time-varying parameters \cite{ahn2013method, chen2012behavioral}, and asymmetric behavioural rules \cite{chen2012behavioral, laval2010mechanism}, have progressively relaxed its deterministic assumptions to accommodate oscillation growth and hysteresis. In parallel, a separate body of work on mixed traffic has produced multi-criteria leader-follower identification frameworks \cite{anand2019calibration, anil2022calibrating, kulkarni2025leader} and behavioural models that represent lateral interactions and heterogeneous vehicle compositions.
However, these two streams of research have developed in near-complete isolation. Newell's extensions were each formulated and validated exclusively in lane-disciplined settings, where the trajectory-translation hypothesis faces none of the complications introduced by lateral manoeuvring, staggered following, and inter-driver heterogeneity. Conversely, the multi-criteria identification frameworks (Section \ref{sec:lf identification}) developed for disordered traffic have served primarily as precursors to calibrating richer behavioural models rather than as tools for evaluating simpler formulations. Whether the behavioural principles embedded in Newell's model retain explanatory power in lane-free conditions, and whether targeted extensions can recover performance without sacrificing parsimony, thus remain open empirical questions.

A further consideration shapes how that evaluation must be conducted. Car-following models differ fundamentally in their intended domain of applicability. Full-spectrum formulations (\cite{treiber2000congested}) are designed to reproduce the complete range of longitudinal driving behaviour, spanning free-flow conditions, the approach phase, constrained following, and emergency braking. Newell's model, by contrast, is explicitly scoped to the core car-following regime. Applying it to trajectories that span transitional segments conflates regimes it was never intended to represent, and risks attributing to theoretical limitations what may instead reflect mismatched scope. A valid empirical test, therefore, requires that core car-following episodes be identified and separated from transitional behaviour prior to any conformity assessment.

The remainder of the paper is organized as follows. Section \ref{sec:3} describes the study site, data collection, and trajectory extraction. Section \ref{sec:4} presents the overall methodology. Section \ref{sec:5} evaluates the validity of the speed-spacing relationship implied by Newell's model. Section \ref{sec:6} introduces a trajectory-shifting framework to directly examine the trajectory-translation hypothesis. Section \ref{sec:7} investigates systematic deviations from the classical formulation and proposes an extension accounting for interaction-boundary effects. Section \ref{sec:8} compares the modelling approaches and summarizes the main findings. Sections \ref{sec:9} and \ref{sec:10} discuss the implications, limitations and future directions of the study, respectively.

\section{Data collection and preliminary analysis} \label{sec:3}
The following section outlines the data collection methodology. This study requires high-resolution trajectory data to accurately analyze vehicle pairs. In such a scenario, video data collection using stationary cameras was considered but found unsuitable due to occlusion of smaller vehicles by larger ones, perspective distortion at oblique recording angles, and limited spatial coverage. Unmanned Aerial Vehicles (UAVs), on the other hand, provide an unobstructed bird's-eye view and were therefore adopted for data collection. However, they have their own limitations, such as battery-limited recording duration, wind-induced video instability, and the costs and regulatory approval involved. A high-fidelity dataset of traffic trajectories was collected using UAVs, equipped with 4K cameras operating at 25 fps, on an urban highway.

\subsection{Study site and data acquisition}
A preliminary site-selection survey was conducted, and several candidate locations were identified. Candidate sites were evaluated on the basis of traffic volume, road geometry, and conditions favourable for reliable video extraction, specifically, straight alignment, minimal side friction, unobstructed overhead clearance, and absence of tree cover or elevated structures. A 1 km section of the AH-45 highway in Chennai (13°11'36.5"N, 80°11'14.3"E) best satisfied these criteria and was selected. The roadway is a divided four-lane facility with two lanes in each direction.

Data collection was carried out on 19 September 2025, intermittently from 11:00-16:00. Seven UAVs flew simultaneously in each session, each covering a 180 m stretch. Adjacent UAVs were overlapped by 15–20~m to ensure seamless coverage. Cameras were aligned with the road median to maintain a consistent bird’s-eye view, and recording was initiated simultaneously across all UAVs. Each session was limited to 12–14 minutes by battery endurance, and seven flights were conducted in total, yielding approximately 90 minutes of usable footage per UAV. For geo-registration, 30 ground control points (GCPs) were distributed along the study section, with each UAV capturing at least six GCPs, including paired GCPs within overlapping zones; global coordinates of all GCPs were recorded using a GPS device. The present analysis uses only northbound traffic from one UAV flight.

\subsection{Data extraction and filtering}
Although the UAVs were equipped with gimbal stabilization, wind, hovering instability, and propeller-induced vibrations caused residual video drift that can introduce errors in extracted trajectories. Video stabilization was therefore applied prior to trajectory extraction using the Scale-Invariant Feature Transform (SIFT) algorithm \cite{lowe2004distinctive}, which produces feature descriptors that are robust to scale, rotation, and image noise \cite{kumar2025stop}. A MATLAB script aligned each frame to a reference frame (the first frame) by extracting SIFT keypoints and descriptors, matching them across frames, and estimating a projective homography to warp each frame into alignment. Frames with insufficient matches were retained in their original form. The resulting stabilized video sequences reduce unwanted camera motion, yielding smoother footage better suited for accurate trajectory extraction.

Data extraction was performed using SAVETRAX \cite{venthuruthiyil2021trajectory}, a MATLAB-based image processing tool designed for efficient and accurate trajectory reconstruction. This semi-automated video processing tool, although time-consuming and labor-intensive, is user-friendly, delivers highly accurate results, and requires less computational resources. Vehicle trajectories were derived by transforming image coordinates into real-world coordinates through a geo-registration process within the software. The global coordinates of distinct reference features were obtained from GPS, with the WGS 84/ UTM zone 44P coordinate system serving as the geospatial reference framework. The extracted trajectory data were subsequently transformed from UTM to a local reference system, allowing for longitudinal analysis along the direction of flow and lateral positioning across the roadway section.

Vehicle trajectories extracted from the videos were first filtered by vehicle class, and only car trajectories were retained for the present study. Subsequently, leader–follower interaction episodes were identified for pairwise trajectory analysis. A succeeding car was classified as a follower when all of the following conditions were satisfied: (i) the spacing between the two vehicles ($h_s$) remained below 20 m throughout the interaction period, (ii) the vehicles exhibited lateral overlap ($O_l$), defined in Eq. \ref{eq:overlap}, (iii) the interaction lasted for at least $10~s$, and (iv) the distance covered during the interaction more than $10~m$.

Given the dense, stop-and-go character of the study site, a tighter $20~m$ threshold was adopted, as it seemed more appropriate than the $30~m$ threshold commonly reported in the literature. Preliminary observations indicated that approaching and diverging behaviour persisted even at a $20~m$ threshold, and these effects were expected to be more pronounced at $30~m$. The criteria of minimum distance and duration were imposed to exclude stationary pairs and retain only meaningful, sufficiently sustained following episodes. Together, these criteria ensured the selection of stable and representative interactions suitable for analyzing car-following behaviour.

\begin{figure}
    \centering
    \includegraphics[width=4.5in]{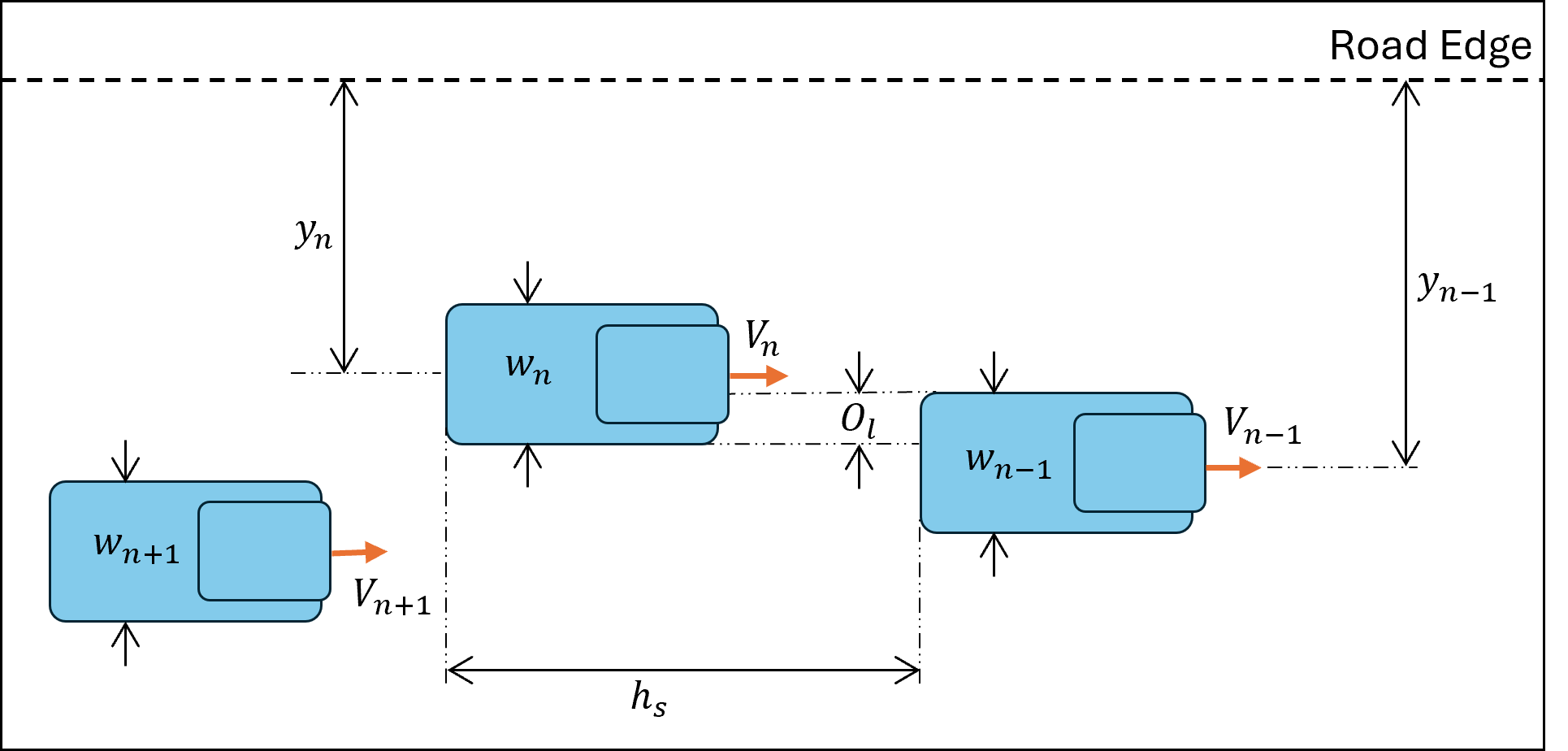}
    \caption{Lateral overlap between leader and follower}
    \label{fig:overlap demo}
\end{figure}

A typical mixed-traffic scenario is illustrated in Fig. \ref{fig:overlap demo}, which provides a geometric interpretation of the overlap measure. In the configuration shown, car $n-1$ (the leader) exhibits lateral overlap with both cars $n$ and $n+1$, each of which lies within 20 m of the leader. Consequently, both criteria (i) and (ii) are satisfied for multiple potential followers simultaneously, yielding an ambiguous follower assignment. To resolve this ambiguity without undue complexity, the present study restricts follower assignment to the nearest vehicle in the opposite direction of movement \cite{papathanasopoulou2018flexible} that satisfies the lateral overlap condition, i.e., car $n$ is taken as the follower of car $n-1$.
\begin{equation}
O_l = \max \left\{ 0, \frac{w_{n-1} + w_n}{2} - \left| y_{n-1} - y_n \right| \right\}
\label{eq:overlap}
\end{equation}
where $w$ denotes vehicle width and $y$ represents the lateral distance from the road edge for the respective vehicles.

Applying these criteria to a 15-minute continuous observation window captured during a single drone flight yielded 72 car–car interactions, which correspond to approximately 88,400 individual data points given the 25 Hz resolution of the data. All subsequent analyses in this study are based on this dataset.

\subsection{Descriptive Statistics}
The selected roadway is a divided four-lane facility, carrying an average traffic volume of approximately 2,200 vehicles per hour per direction. The traffic stream is predominantly heterogeneous, with two-wheelers (40\%), cars (24\%), and heavy vehicles (20\%) accounting for the major share.

\begin{table}[htbp]
\centering
\caption{Descriptive statistics of car-following pairs ($N = 74$)}
\label{tab:descriptive_statistics}
\begin{tabular}{p{6cm} p{3cm} p{5cm}}
\hline
\rowcolor{blue!20}
\textbf{Variable} & \textbf{Statistic} & \textbf{Value} \\
\hline

\rowcolor{blue!10}
\multicolumn{3}{l}{\textbf{Interaction characteristics}} \\

Following duration & Mean (IQR) & 49 s (25.7--66.4 s) \\
Distance covered & Mean (IQR) & 96 m (46.7--146.9 m) \\
Spacing variance & Mean & 6.27 m$^2$ \\
Spacing RMSE & Mean & 2.33 m \\

\hline

\rowcolor{blue!10}
\multicolumn{3}{l}{\textbf{Lateral positioning}} \\

Pair-averaged lateral overlap & Mean (SD) & 1~m (0.31 m) \\
Within-pair overlap variance & Mean & 0.04 m$^2$ \\
Between-pair variance in mean overlap & --- & 0.1 m$^2$ \\

\hline

\rowcolor{blue!10}
\multicolumn{3}{l}{\textbf{Longitudinal dynamics}} \\

Pair-averaged speed & Mean & 2.2 m/s ($\approx$7.9 km/h) \\
Maximum speed (per pair) & Mean & 7 m/s ($\approx$25.2 km/h) \\
Within-pair speed variance & Mean & 4.95 m$^2$/s$^2$ \\
Between-pair variance in mean speed & --- & 1.72 m$^2$/s$^2$ \\

\hline
\end{tabular}

\vspace{0.2cm}
\footnotesize{IQR = interquartile range; SD = standard deviation; RMSE = root mean square error}

\end{table}

Table \ref{tab:descriptive_statistics} presents the descriptive statistics for this filtered sample, organized by interaction characteristics, lateral positioning, and longitudinal dynamics. The wide interquartile ranges in the following duration and distance covered point to substantial heterogeneity in interaction length, reflecting the varied spacing and speed conditions across the sample. The mean pair-averaged lateral overlap of $\approx$ 1~m indicates that followers maintain a substantial lateral overlap with their leaders throughout the interaction. Low within-pair overlap variance (0.04 m²) alongside modest between-pair variance (0.1 m²) indicates that partial lateral alignment is a stable feature of the car–car following regime rather than an artefact of isolated pairs. The low mean pair-averaged speed of 2.2 m/s is consistent with congested, near-stop-and-go conditions, while the notably high mean within-pair speed variance (4.95 m²/s²) points to pronounced deceleration–acceleration cycles within individual interactions. The between-pair variance in mean speed (1.72 m²/s²) suggests that pairs were captured at different phases of the stop-and-go wave, resulting in appreciably different mean operating speeds across the sample.

\section{Conceptual Framework: Testing Newell’s Hypothesis}\label{sec:4}
This study evaluates whether the behavioural principles of Newell's model are consistent with empirical observations of lane-free traffic. The analysis is structured as a sequence of complementary tests that examine different implications of the Newell hypothesis.
\begin{figure}[H]
\centering
\begin{subfigure}[b]{0.45\linewidth}
\centering
\includegraphics[width=\linewidth,height=8cm,keepaspectratio]{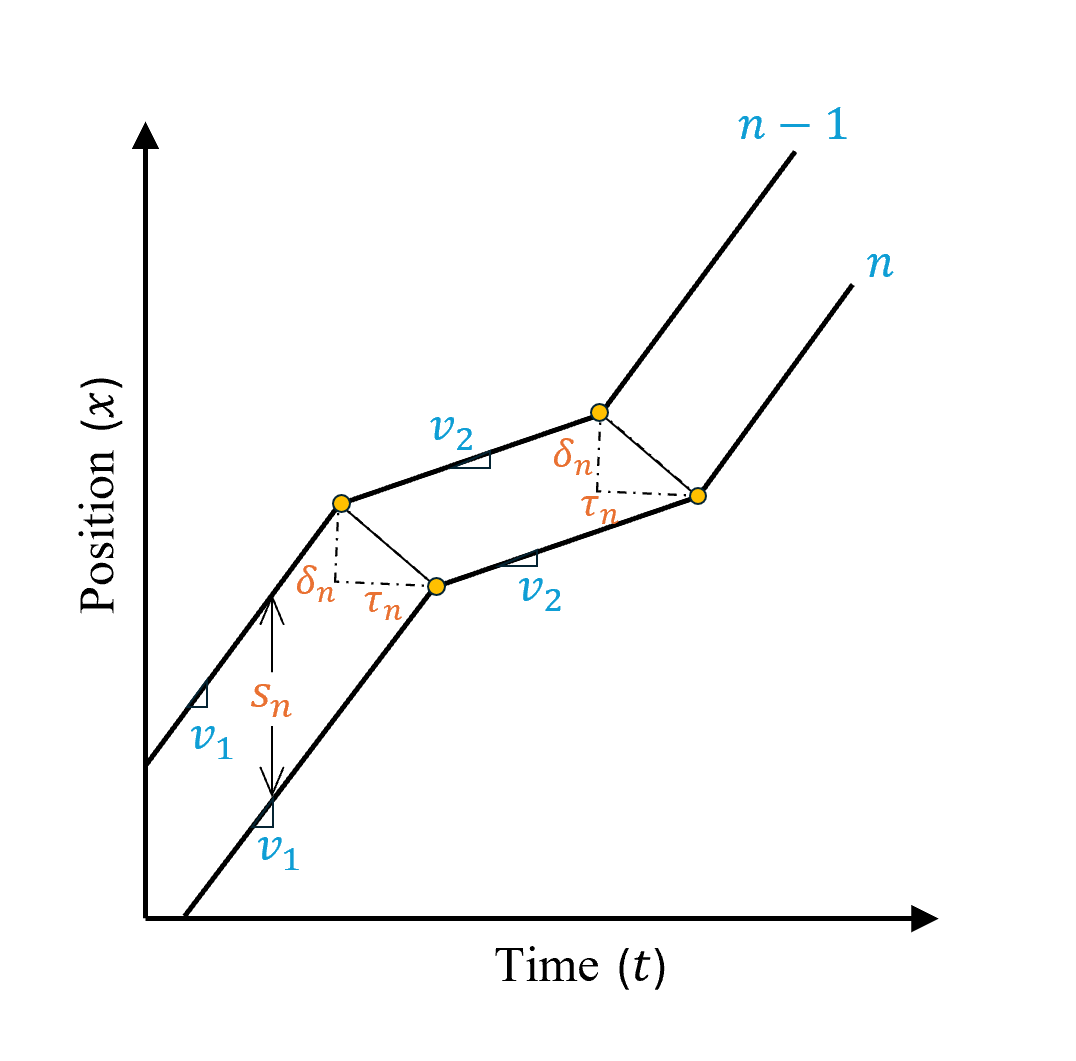}
\caption{}
\end{subfigure}
\hspace{0.02\linewidth}
\begin{subfigure}[b]{0.45\linewidth}
\centering
\includegraphics[width=\linewidth,height=8cm,keepaspectratio]{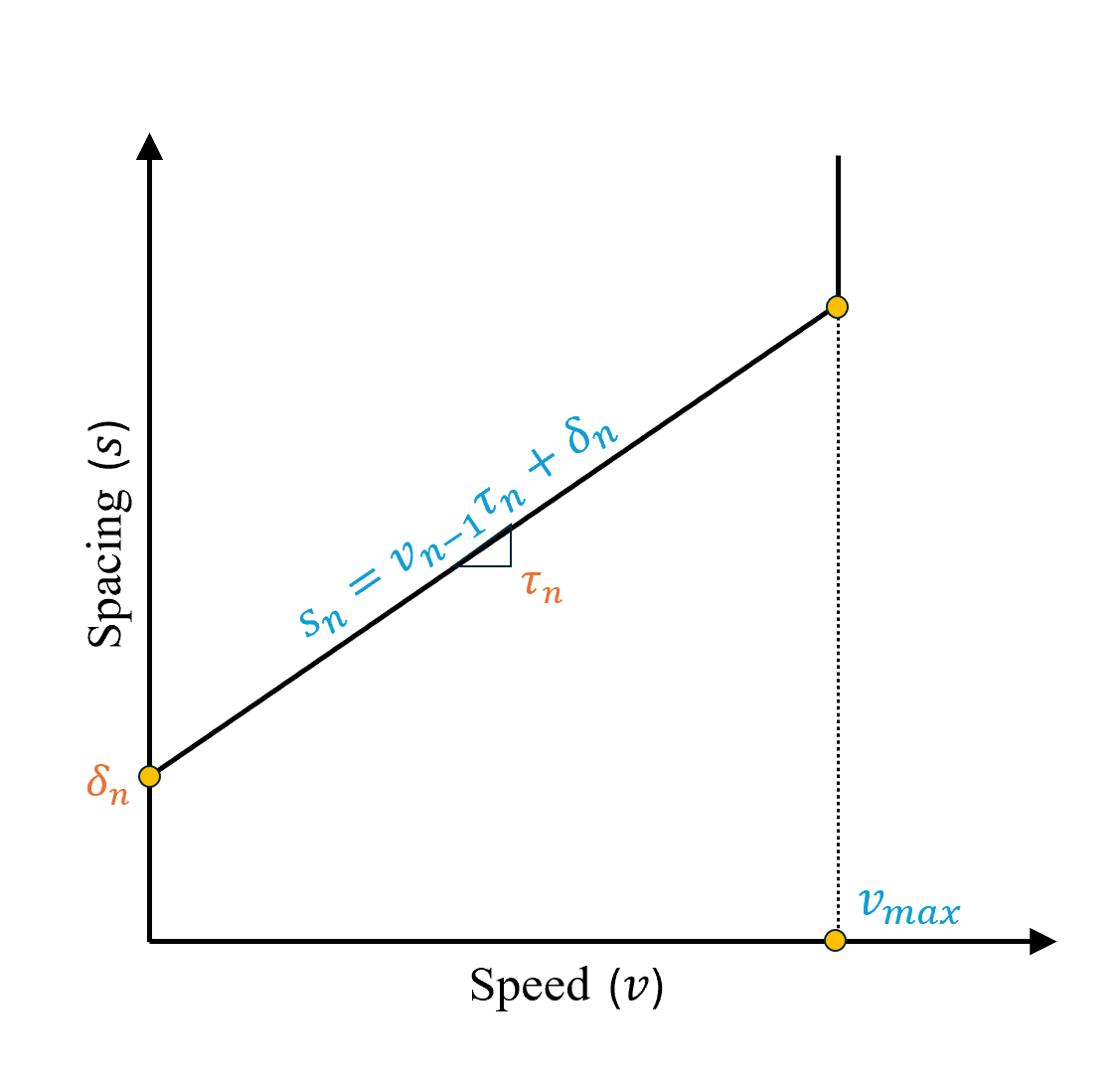}
\caption{}
\end{subfigure}
\caption{(a) Trajectory translation, (b) Speed-spacing relationship}
\label{fig:newell}
\end{figure}
\noindent The first implication of Newell's model concerns a linear speed–-spacing relationship in congested traffic, the validity of which constitutes a direct test of the model. To assess the extent to which this relationship is supported by the observed data, regression-based approaches were used to estimate the corresponding parameters. Supplementary covariates, namely acceleration and deceleration, are subsequently introduced to assess whether they improve explanatory power. In addition, the analysis quantifies the extent to which behavioural heterogeneity between vehicle pairs affects the stability of the estimated parameters.

The second implication concerns the spatio-temporal organization of vehicle trajectories. Under the model, the follower's trajectory should be a rigid translation of the leader's in the space–time $(x-t)$ plane. This proposition was examined by shifting the leader's trajectory by appropriate offsets and comparing the resulting path directly with the follower's observed path. The degree of agreement between the two serves as a behavioural test of the translation principle. To accommodate systematic deviations near the boundaries of the interaction interval, an extension is introduced that permits spacing to vary in these regions while preserving the core translation structure.

Like Duret et al.\cite{duret2008estimating} and Chiabut et al.\cite{chiabaut2010heterogeneous}, this study evaluates microscopic vehicle interactions through pair-resolved implementations of Newell’s CF model. The studies differ, however, in their data environments, estimation approaches, and modeling scopes. Both earlier works draw from the NGSIM dataset, which captures lane-disciplined highway congestion to analyze lane-disciplined, structured highway congestion; the present study uses high-resolution UAV footage of lane-free, heterogeneous Indian traffic. Their underlying mathematical frameworks also diverge: Duret et al.\cite{duret2008estimating} contrast linear regression with simulation-based lead-vehicle optimization, while Chiabut et al.\cite{chiabaut2010heterogeneous} track the stationarity of shockwave parameters across leader speed classes. The present study combines aggregate and pair-specific regression with discrete, frame-level trajectory-shifting variance minimization. Most notably, whereas earlier frameworks suffer from high data exclusion rates due to parameter inconsistencies, which may stem from transitional boundary effects, our study addresses this limitation directly by introducing a novel piecewise quadratic boundary correction to capture these effects. Table \ref{tab:comparison_previous} summarizes the comparison among the three studies.

\begin{table*}[ht]
\centering
\caption{Comparison of the present study with previous similar works}
\label{tab:comparison_previous}
\renewcommand{\arraystretch}{1.4}

\begin{tabularx}{\textwidth}{p{2.2cm} p{3.2cm} p{3.2cm} X}
\hline
\rowcolor{blue!20}
\textbf{Aspect} &
\textbf{Duret et al. (2008)} \cite{duret2008estimating} &
\textbf{Chiabaut et al. (2010)} \cite{chiabaut2010heterogeneous} &
\textbf{This study} \\
\hline

\rowcolor{blue!8}
Primary objective &
Quantify individual speed-spacing relationships &
Connect microscopic parameters with macroscopic patterns &
Evaluate the validity of Newell's model while improving the estimation through extended approaches \\

Boundary transitions &
Not considered &
Not considered &
Identified, quantified, and corrected via a piecewise quadratic extension \\

\rowcolor{blue!8}
Cross-method comparison &
No &
No &
Regression and trajectory-shifting approaches compared directly \\

Error analysis &
RMSE of simulated trajectories &
Stationarity of estimated parameters &
Estimation bias quantified before and after boundary correction \\

\rowcolor{blue!8}
Parameter specification &
Two fixed parameters per driver: wave speed and minimum jam spacing &
Two fixed parameters per driver: jam spacing and time offset &
Two pair-specific parameters; multiple regression adds acceleration covariates; piecewise extension adds four boundary parameters \\

\hline
\end{tabularx}
\end{table*}

\section{Speed-Spacing Relationship: Testing the Linear Hypothesis}\label{sec:5}
\subsection{Basic Aggregate Linear Regression Model} \label{sec:basic agg}
A central implication of Newell’s car-following model is the existence of a linear relationship between the spacing maintained by a follower and speed under car-following conditions, i.e., the spacing between the two vehicles should increase proportionally with the leader’s speed (Fig. \ref{fig:newell}). This linearity is what makes Newell's model elegant. The entire congested regime reduces to a two-parameter straight line. The relationship can be derived by substituting $x_n{(t)}$ from Eq.1 in spacing $s_n(t)$:
\begin{equation}
\begin{aligned}
s_n(t) &= x_{n-1}(t) - x_n(t) \\
       &= x_{n-1}(t) - x_{n-1}(t - \tau_n) + \delta_n
\end{aligned}
\label{eq:spacing1}
\end{equation}

\noindent For steady-state following the distance traveled by the leader during $\tau_n$ is given by:
\begin{equation}
    x_{n-1}(t) - x_{n-1}(t - \tau_n) = v_{n-1} \cdot \tau_n
    \label{eq:spacing2}
\end{equation}

\noindent Therefore using Eq.\ref{eq:spacing2} in Eq.\ref{eq:spacing1}:
\begin{equation}
    \boxed{s_n = v_{n-1} \cdot \tau_n + \delta_n}
    \label{eq:linear relationship}
\end{equation}

\noindent Under Newell's formulation, both $\tau$ and $\delta$ are theoretically expected to remain constant for a given driver throughout a CF episode, implying the linear speed–spacing relationship given in Eq. \ref{eq:linear relationship}. Linear regression provides a natural framework for assessing whether this relationship is consistent with observed data. As an initial reference benchmark, an aggregate model was estimated by pooling all observations across the identified LF pairs, thereby imposing inter-driver homogeneity \cite{jabari2014probabilistic}, that is, a single parameter set ($\tau$, $\delta$) is assumed to govern spacing behaviour across the full sample. Although this constitutes a deliberate simplification of the underlying heterogeneity in driver behaviour, the aggregate model establishes a useful baseline against which pair-level estimates can subsequently be evaluated. Spacing ($s$) was specified as the dependent variable and leader speed ($v$) as the sole explanatory variable, yielding the linear form:

\begin{equation}
   s_i = \beta_0 + \beta_v \cdot v_i + \varepsilon_i
   \label{eq:agg vel}
\end{equation}
In this formulation, the intercept $\beta_0$ and slope $\beta_v$ serve as the representations of $\delta$ and $\tau$, respectively, while $\epsilon_i$ is the stochastic error term. It should be noted that the validity of standard inferential tests is contingent on the approximate normality and statistical independence of the residuals; where this assumption does not hold, the least-squares estimates can still be treated as descriptors of the underlying patterns and interpreted accordingly.

\begin{figure}[H]
    \centering
    \includegraphics[width=0.7\linewidth]{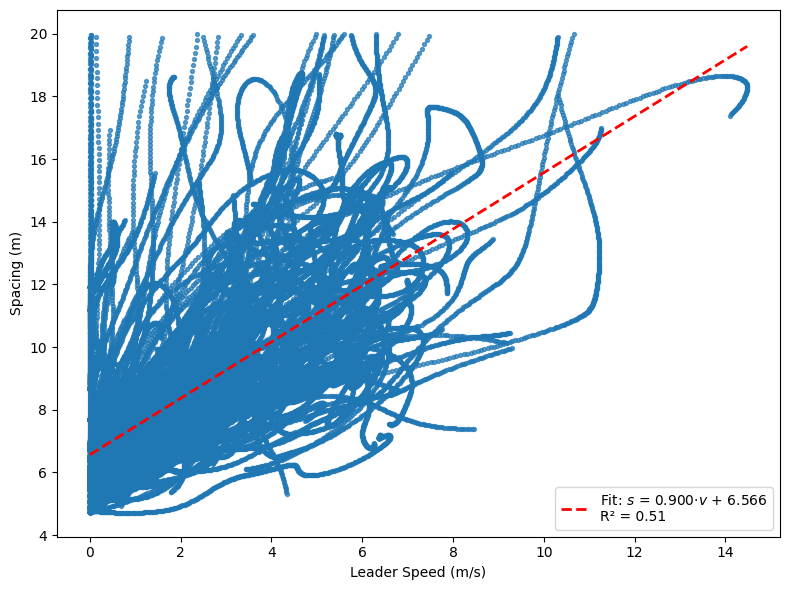}
    \caption{Linear regression on aggregate data}
    \label{fig:simple LR}
\end{figure}
\noindent The results (Fig.~\ref{fig:simple LR}) indicate that the aggregate linear model explains nearly half of the observed variation in spacing ($R^2=0.54$) with $\tau=0.92s$ and $\delta=6.4m$. It means that although a positive speed–spacing relationship exists in the data, a substantial portion of the observed variability cannot be explained by a homogeneous response to speed.

One possible source of scatter is that followers are not always in a steady-state car-following regime. During portions of an interaction, drivers accelerate toward the leader or actively adjust their spacing, conditions under which the equilibrium relationship assumed by the model does not strictly hold.
In the following section, the analysis is therefore extended to account for the non-steady-state portions of car-following interactions, incorporating acceleration/deceleration as an additional explanatory variable(s), with the expectation that this inclusion will improve model fit.

\subsection{Aggregate Model with Acceleration and Deceleration} \label{sec:agg acc dec}
In the aggregate specification presented in the previous section, spacing is determined solely by the leader's speed, with the assumption that driver behaviour is symmetric with respect to changes in speed and that spacing adjusts instantaneously to its equilibrium value. However, empirical observations of car-following behaviour suggest that these assumptions may not hold in practice. According to some studies \cite{newell1962theories, treiterer1974hysteresis,zhang1999mathematical}, drivers may exhibit different responses during acceleration and deceleration phases, leading to asymmetric spacing behaviour at similar speeds. Thus, spacing is influenced not only by speed but also by the dynamic state of motion.
To examine these effects and to improve the explanatory power of the aggregate regression model, acceleration-related variables were incorporated. Two specifications were tested. In the first specification, acceleration and deceleration were combined into a single variable, with deceleration represented as negative values. In the second specification, acceleration and deceleration were introduced as distinct explanatory terms to capture potential asymmetries in driver responses. The extended specifications can be expressed as:
\begin{equation}
s_i = \beta_0 + \beta_v v_i + \beta_a\, a_i + \varepsilon_i
\label{eq:agg acc}
\end{equation}
\begin{equation}
s_i = \beta_0 + \beta_v v_i + \beta_a\, a_i^+ + \beta_d\, a_i^- + \varepsilon_i
\label{eq:agg acc dec}
\end{equation}

\noindent where $a_i$ represents acceleration of the leader, $a_i^+=max(0, a_i)$ and $a_i^-=max(-a_i,0)$ i.e., deceleration is treated as a distinct variable rather than a signed value. $\beta_a$ and $\beta_d$ are parameters capturing their respective effects on spacing. This formulation allows the model to distinguish between situations in which a driver is accelerating and those in which the driver is decelerating, even if the instantaneous speed is the same.
If such asymmetric behaviour is present in the data, the extended model should exhibit a noticeable improvement in explanatory power relative to the baseline specification. The results, however, indicate that including acceleration and deceleration does not yield a significant improvement in model performance. The $R^2$ increases slightly to approximately $0.55$ in both the first (Eq. \ref{eq:agg acc}) and second (Eq. \ref{eq:agg acc dec}) specifications.

This finding does not merely reflect noise in the data; rather, it suggests that acceleration and deceleration contribute little to explaining variability in spacing at the aggregate level. Consequently, accounting for acceleration and deceleration alone is insufficient to capture the observed variability in spacing, and a framework that explicitly models differences across individual leader–follower pairs appears necessary. This motivated the development of pair-specific models, examined in the subsequent section.
\subsection{Driver Heterogeneity: Pair-Specific Model}
The aggregate models in Sections \ref{sec:basic agg} and \ref{sec:agg acc dec} leave substantial spacing variability unexplained. A plausible source is the inherent heterogeneity of driving behaviour. Individual drivers differ in risk perception, reaction time, and vehicle performance, each of which influences the spacing maintained at a given speed. When heterogeneous observations are pooled into a single regression, the estimated coefficients reflect the average relationship across the sample, not the relationship for any individual pair. A single global speed–spacing function may therefore not hold uniformly across all pairs. To capture this heterogeneity, a fixed-effects regression model was estimated via the least-squares dummy variable (LSDV) approach, allowing the intercept and slope of the speed–spacing relationship to vary freely across leader–follower pairs. Four specifications were considered.

As a baseline reference, a null specification retaining only the pair-specific intercept was estimated:
\begin{equation}
    s_{i,k} = \alpha_{k}+\varepsilon_{i,k}
    \label{eq:null model}
\end{equation}
where  $s_{i,k}$ denotes the spacing for observation $i$ belonging to LF pair $k$, $\alpha_{k}$ is the pair-specific intercept. This benchmark isolates the explanatory contribution of each subsequent covariate.

The first substantive specification introduces speed as the sole explanatory variable, directly complementing the linear relationship in Eq. \ref{eq:agg vel}:
\begin{equation}
  s_{i,k} = \alpha_k + \beta_{1k} v_{i,k} + \varepsilon_{i,k}
\end{equation}
where $v_{i,k}$ is the leader speed and $\beta_{1k}$ is the pair-specific speed sensitivity coefficient.

A specification incorporating a single acceleration term and is complementary to Eq. \ref{eq:agg acc}, was then estimated:
\begin{equation}
  s_{i,k} = \alpha_k + \beta_{1k} v_{i,k} + \beta_{2k} a_{i,k} + \varepsilon_{i,k}
\end{equation}
where $a_{i,k}$ is the signed acceleration of follower and $\beta_{2k}$ is the pair-specific acceleration response coefficient.

Finally, acceleration is disaggregated into its positive and negative components, complementing the asymmetric formulation in  Eq. \ref{eq:agg acc dec}. It allows the model to capture asymmetric spacing responses to accelerating and decelerating manoeuvres.
\begin{equation}
  s_{i,k} = \alpha_k + \beta_{1k} v_{i,k} + \beta_{2k} a^{+}_{i,k} + \beta_{3k} a^{-}_{i,k} + \varepsilon_{i,k}
\end{equation}
where $a^{+}_{i,k} = \max(a_{i,k},\,0)$ and $a^{-}_{i,k} = \min(a_{i,k},\,0)$, and $\beta_{2k}$, $\beta_{3k}$ are the corresponding pair-specific coefficients.

Across all specifications, the pair-specific parameters are expressed through an LSDV structure. If $K$ denotes the total number of LF pairs, then:

\[
\begin{aligned}
\alpha_k &= \alpha_0 + \sum_{j=1}^{K} \delta_{0j}\, \mathbb{I}_{j},
\qquad \forall\, k \in \{1,\dots,K\}
\\[8pt]
\beta_{mk} &= \beta_m + \sum_{j=1}^{K} \delta_{mj}\, \mathbb{I}_{j},
\qquad \forall\, k \in \{1,\dots,K\},\; \forall\, m \in \{1,2,3\}
\end{aligned}
\]

\noindent where $\alpha_0$ and $\beta_m$ are the baseline intercept and slope coefficients of the reference pair, $\delta_{0j}$ and $\delta_{mj}$ denote the intercept and slope deviations of pair $j$ from the respective baseline values, and $\mathbb{I}_j$ is a binary indicator equal to 1 if the observation belongs to pair $j$ and 0 otherwise. The pair-specific parameters $\alpha_k$ and $\beta_{mk}$ thereby absorb all between-pair heterogeneity.

The pair-specific fixed effects yield a marked improvement in model performance. The null model (Eq. \ref{eq:null model}) achieves a pooled $R^2=0.22$, quantifying variance attributable to pair identity alone. The substantial gap between this benchmark and the subsequent specifications underscores the indispensability of speed and acceleration as explanatory variables. Introducing speed raises the pooled $R^2= 0.78$, substantially exceeding any aggregate specification. Augmenting the model with a combined acceleration term yields a further improvement to $R^2 = 0.83$, while the disaggregated formulation achieves $R^2 = 0.84$. Like the aggregate model, $\alpha_m$ and $\beta_{1k}$ represent $\delta$ and $\tau$, respectively. The mean estimates of $\delta$ and $\tau$ are $\approx$6.6~m and $\approx$1.2~s, which are consistent with those obtained from the aggregate model.

This incremental pattern suggests that speed accounts for the bulk of explainable variance, with acceleration effects contributing a smaller but meaningful share. More importantly, the variability observed in aggregate models reflects systematic, structured differences among drivers rather than random noise, a pattern broadly consistent with the aggressive–cautious continuum documented in the car-following literature. From the standpoint of Newell's model, this implies that $\tau$ and $\delta$ are pair-specific quantities rather than universal constants. A single aggregate calibration risks absorbing systematic behavioural heterogeneity into residual error and misrepresenting the underlying process.

The above findings establish heterogeneity as a dominant structural factor governing spacing behaviour. By explicitly accommodating pair-specific effects, the regression framework recovers a strong linear speed–spacing relationship consistent with Newell's theoretical architecture, while simultaneously revealing the considerable extent to which the governing parameters vary across the traffic stream. This finding raises a natural question: whether the individualized behaviour identified through regression is also embedded in the spatio-temporal structure of vehicle trajectories? This is examined in the subsequent section through the trajectory-shifting framework.
\section{Trajectory-Shifting Framework: Testing Spatiotemporal Similarity of Trajectories} \label{sec: traj shifting} \label{sec:6}
The regression-based analyses presented in the previous section offer indirect, statistically mediated evidence for Newell's CF model. A more direct test is possible by exploiting the central structural assumption of the model: the follower's trajectory is a spatiotemporal translation of the leader's, displaced by fixed offsets $(\tau, \delta)$ in time and space. The trajectory-shifting method operationalizes this directly, estimating $(\hat\tau, \hat\delta)$ as the shifts required to align the two observed trajectories, without imposing any predefined functional relationship between kinematic variables. Unlike regression, which tests the model indirectly in the speed–spacing plane, this approach evaluates the translation principle in the space-time domain on its own terms and yields pair-specific parameter estimates without pooling across drivers.

The rationale for the method lies in the asymmetry (temporal mismatch) inherent in car-following behaviour. In its unshifted form, the inter-vehicle spacing observed at time $t$ conflates two kinematically distinct regimes-- steady-state following, during which both vehicles travel at comparable speeds, and their trajectories are approximately parallel; and transitional regimes, during which the follower responds to a prior leader manoeuvre. In the latter, the trajectories exhibit diverging or converging slopes, corresponding to differences in instantaneous speed, which generate pronounced fluctuations in observed spacing. Critically, these fluctuations do not reflect an intrinsic driving behaviour property but they arise as an artefact of temporal misalignment. The follower at time $t$ responds not to the leader's contemporaneous state but to the leader's state at an earlier instant ($t-\hat{\tau}$). Shifting the leader's trajectory forward by $\hat\tau$ brings this earlier state into temporal correspondence with the follower's present response, rendering the two trajectories locally parallel across all regimes and driving the residual spacing towards a constant value, the jam spacing $\delta$. The optimal time shift $\hat\tau$ is therefore identified as the value that minimizes the variance of the shifted spacing series (Eq. \ref{eq:15}).

Let $x^l(t)$ and $x^f(t)$ denote the positions of the leader and follower, respectively. For a given candidate time lag $\tau$, the spacing is defined as
\begin{equation}
s_{\tau}(t) = x^l(t - \tau) - x^f(t)
\label{eq:15}
\end{equation}
The variance of $s_{\tau}(t)$ over the observation interval is then computed, and the optimal time lag $\hat\tau$ is obtained by solving
\begin{equation}
\hat{\tau} = \arg\min_{\tau} \, \mathrm{Var}\big(s_{\tau}(t)\big)
\end{equation}
Once the optimal time lag is identified, the corresponding spatial offset $\hat{\delta}$ is calculated as the mean of the spacing values associated with $\hat{\tau}$:
\begin{equation}
\hat{\delta} = \mathbb{E}\big[s_{\hat{\tau}}(t)\big]
\end{equation}
Due to the discrete nature of video data, the optimization procedure is implemented using incremental time shifts at the frame level. For each LF pair, the leader's trajectory is shifted forward in time in discrete increments of $\Delta t = 1/25$ s, given the sampling frequency of 25 Hz. At each candidate shift $\tau=s\,\Delta t$, series length is held constant by prepending $s$ historical leader data points at the start and discarding $s$ points from the tail. The pair $(\hat{\tau}, \hat{\delta})$ minimizing the variability of the resulting spacing series constitutes the estimated parameters. The corresponding shifted leader trajectory, $x_l(t-\hat{\tau})-\hat{\delta}$, serves as the predicted trajectory of the follower.

The performance of the method is assessed by comparing the reconstructed follower trajectory against the observed one using two complementary quantitative measures: residual spacing variance and RMSE. To facilitate cross-pairs comparison given different baseline dynamics, a normalized performance indicator, denoted $\bar{R^2}$, is introduced. This measure quantifies the proportional reduction in spacing variance attributable to the temporal shift of the leader's trajectory, relative to the unshifted baseline.

\begin{equation}
\bar{R}^2_{k}
=
1 - 
\frac{
\mathrm{Var}_{t \in T}\!\left( x_{k}^l(t - \hat{\tau}) - x_{k}^f(t) \right)
}{
\mathrm{Var}_{t \in T}\!\left( x_{k}^l(t) - x_{k}^f(t) \right)
}
\end{equation}
where $\bar{R}_k^{2}$ is the variance-reduction index for pair $k$; $x_k^l(t)$ and $x_k^f(t)$ denote the longitudinal positions of the leader and follower vehicles of pair $k$ at time $t$, respectively. $x_k^l(t - \hat{\tau})$ is the leader trajectory shifted in time by $\hat{\tau}$; $T$ is the trimmed time domain common to both the shifted and unshifted series, and $\mathrm{Var}_{t \in T}(\cdot)$ denotes the sample variance computed over $T$. Analogously to $R^2$ in regression, $\bar{R}_k^2 \rightarrow 1$ indicates that the trajectory translation accounts for nearly all observed spacing variability, whereas $\bar{R}_k^2 \rightarrow 0$ implies that temporal realignment yields little improvement over the unshifted baseline.

\begin{figure}[H]
\centering
\begin{subfigure}[b]{0.45\linewidth}
  \centering
  \includegraphics[width=\linewidth,height=4.5cm,keepaspectratio]{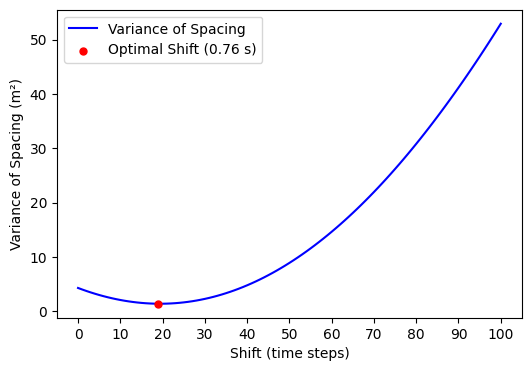}
  \caption{Variation of variance with shift}
\end{subfigure}\hfill
\begin{subfigure}[b]{0.45\linewidth}
  \centering
  \includegraphics[width=\linewidth,height=4.5cm,keepaspectratio]{\detokenize{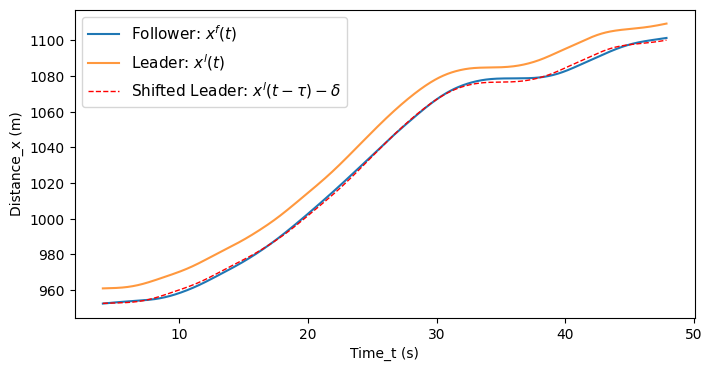}}
  \caption{Observed and shifted follower}
\end{subfigure}
\caption{ Trajectory shifting}
\label{fig:shift}
\end{figure}
\noindent The method yields a reasonably good approximation of follower behaviour across the dataset. The average $\bar{R}^{2}$ is approximately $0.66$, and the average spacing RMSE decreases from an initial value of $2.33\,\mathrm{m}$ to $1.2\,\mathrm{m}$ following the optimal shift. These figures suggest that the translated trajectory captures a substantial portion of the observed spacing dynamics. For a number of pairs, the shifted leader trajectory closely overlaps with the observed follower trajectory (Fig.~\ref{fig:shift}), reflecting strong behavioural consistency with Newell's constant-shift assumption.

\begin{figure}
    \centering
    \includegraphics[width=0.5\linewidth]{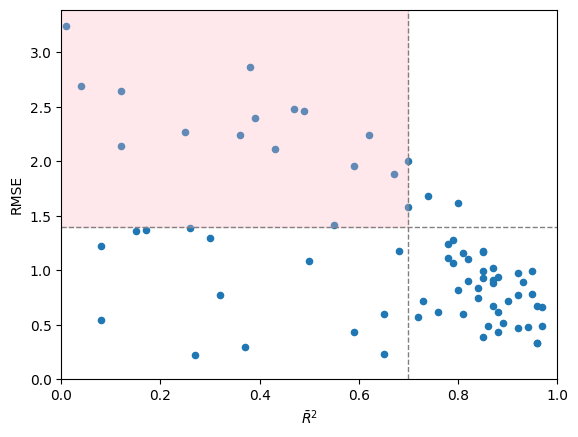}
    \caption{$\bar{R^2}$ vs RMSE}
    \label{fig:R2 vs RMSE}
\end{figure}

However, the agreement is not uniform across LF pairs. For a substantial subset, systematic deviations between the predicted and observed trajectories are evident. Pairs satisfying $\bar{R^2}<0.7$ and $RMSE>1.4\, \mathrm{m}$ (Fig. \ref{fig:R2 vs RMSE}) are classified as exhibiting substantial deviations. Such pairs account for approximately $20\%$ of the dataset, a non-trivial share that warrants closer examination. Visual inspection of these pairs shows that in 12 of 15 cases, the mismatch concentrates near the boundaries of the interaction interval, i.e., at the onset and termination of the car-following episode. There, the constant-spacing assumption underlying the variance-minimization criterion breaks down, since the optimization procedure averages across qualitatively distinct behavioural regimes. More broadly, pairs with high agreement typically correspond to sustained periods of steady following, whereas pairs with poor fit are characterized by pronounced transitional phases at the onset and termination of the interaction (Fig. \ref{fig:ee main}), or sometimes abrupt mid-interaction speed changes. These observations indicate that residual deviations from Newell's model are not random but are systematically concentrated at the entry and exit phases of CF interactions. The nature and extent of these boundary effects are examined in detail in the following section.

\section{Understanding Deviations: Edge Effects} \label{sec:7}
\begin{figure}[htbp]
\centering

\begin{subfigure}[b]{0.45\textwidth}
    \centering
    \includegraphics[width=\textwidth]{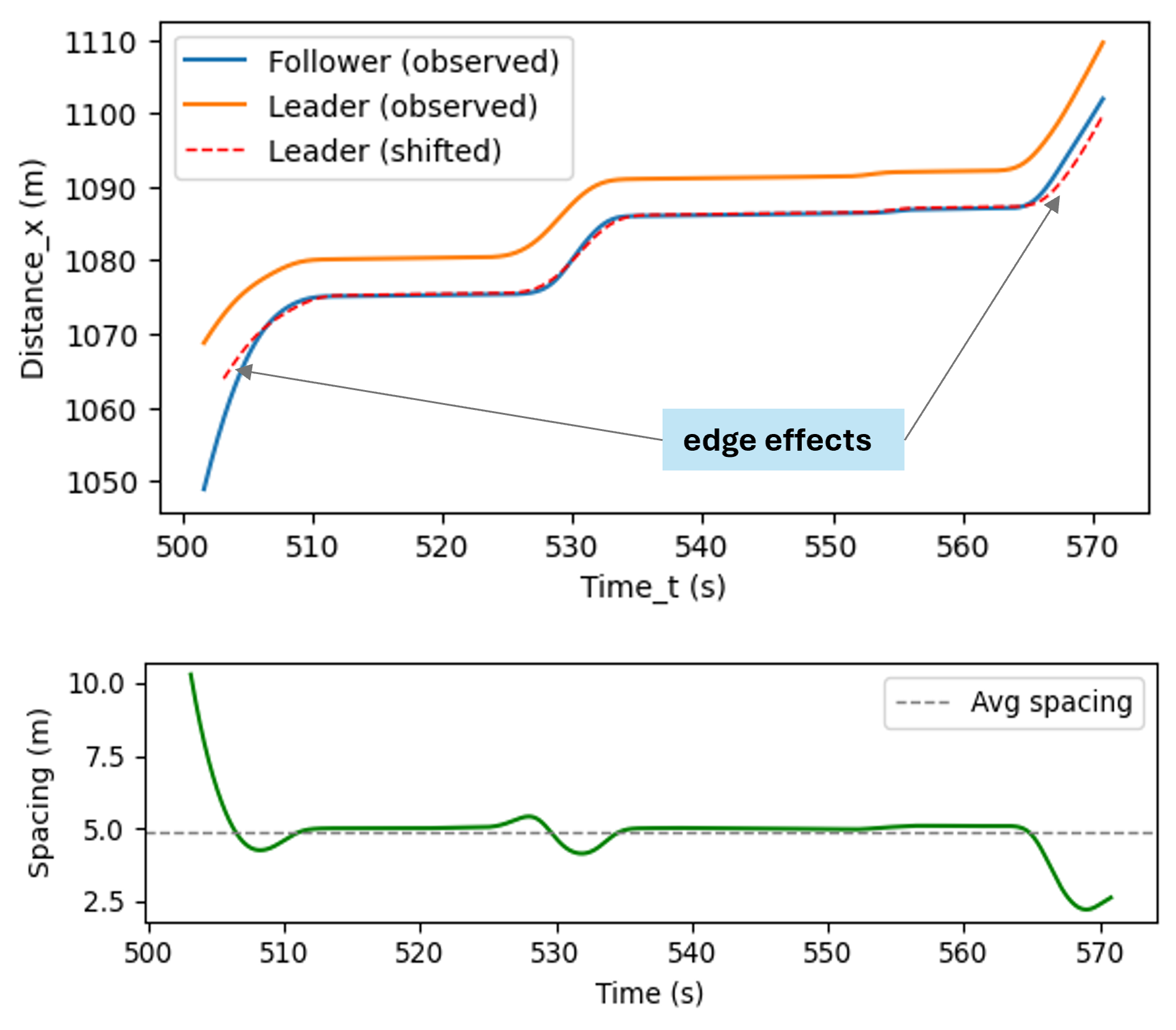}
    \caption{Merging and diverging edge effects}
    \label{fig:ee1}
\end{subfigure}
\hfill
\begin{subfigure}[b]{0.45\textwidth}
    \centering
    \includegraphics[width=\textwidth]{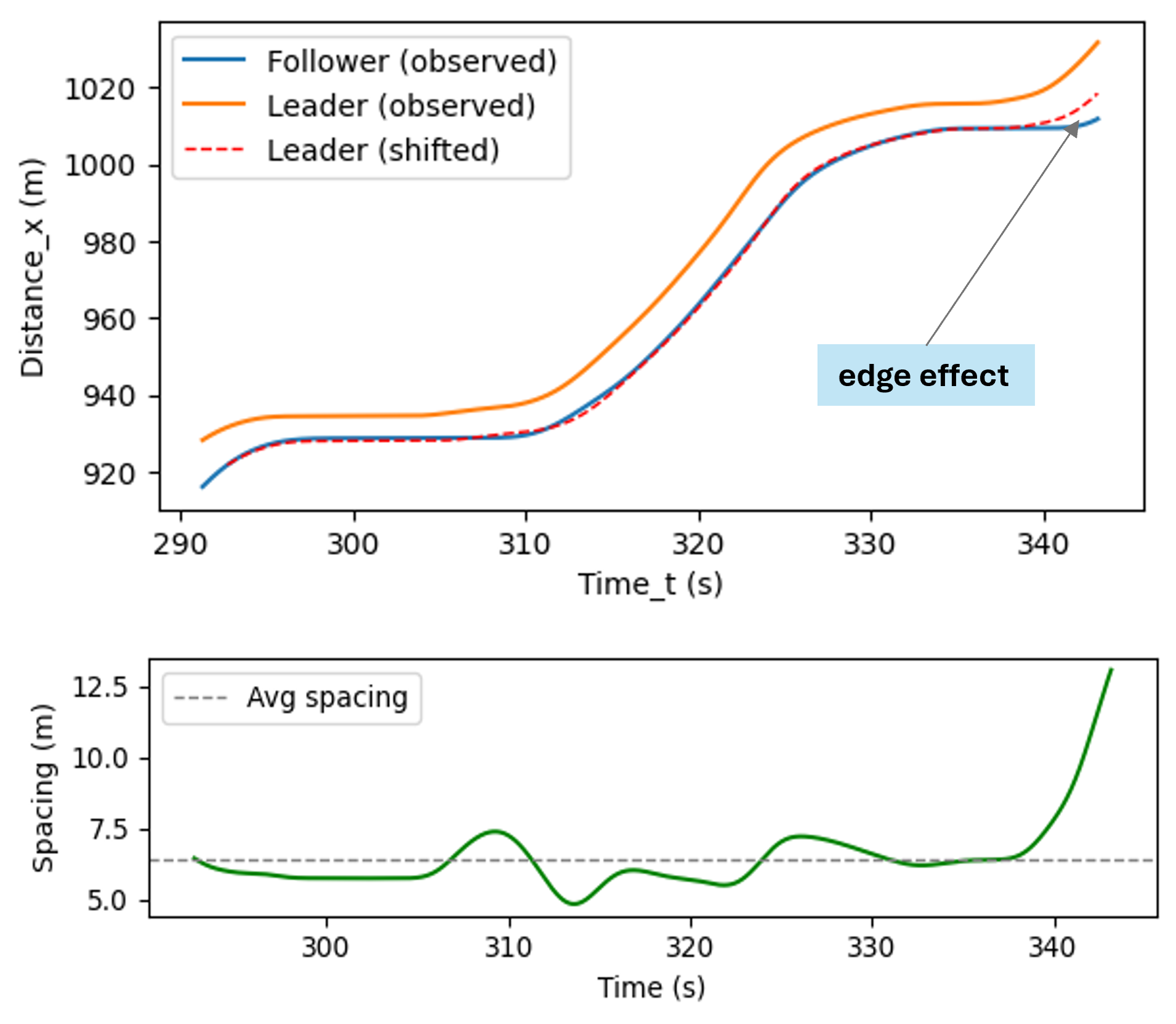}
    \caption{Diverging edge effects}
    \label{fig:ee2}
\end{subfigure}
\caption{Evidence of edge effects in dataset}
\label{fig:ee main}
\end{figure}

As noted above, the discrepancies are concentrated predominantly at the onset and termination of the car-following episode, corresponding to regime transitions in which the follower is either closing in or pulling away from the leader. During such periods, the steady-state assumption implicit in the basic Newell formulation no longer holds. The behaviour in these phases is inherently non-stationary and cannot be captured by a single constant time–space translation, i.e., a single pair of $\tau$ and $\delta$. Empirical data confirms this. When the spacing between the shifted leader and the observed follower trajectory is examined over time, spacing variation remains relatively low during the central portion of the interaction but increases markedly near the beginning and/or end (Fig. \ref{fig:ee main}). From a behavioural perspective, a car-following interaction consists of at least two distinct regimes: a core CF regime, in which the follower maintains approximately constant spacing relative to the leader, and a transition regime, in which spacing evolves dynamically as the follower adjusts to changing conditions. The Newell's model is intended to describe only the former. Regime transitions may involve stronger acceleration or deceleration than what the basic model implicitly assumes. As a result, estimating a single set of parameters across the entire trajectory can yield biased or unstable estimates. The optimization procedure (Section \ref{sec: traj shifting}) attempts to compromise between steady-state following and transition segments, producing parameter values that accurately represent neither.Thus, including edge-effect segments in the estimation process may obscure the underlying structure of steady-state car-following behaviour. Consequently, an accurate assessment of the Newell model requires precise isolation of the core car-following portions of the trajectory and the explicit extension of the model to account for regime-transition dynamics.

\noindent These transitional segments are referred to hereafter as edge-effects (EE). The following subsection characterises them and evaluates whether the Newell model's limitations reflect a fundamental failure of its assumptions or merely the inclusion of behavioural regimes beyond its intended scope.

\subsection{Isolating Regime Transitions: Extended Piecewise Model}
\begin{figure}
    \centering
    \includegraphics[width=1\linewidth]{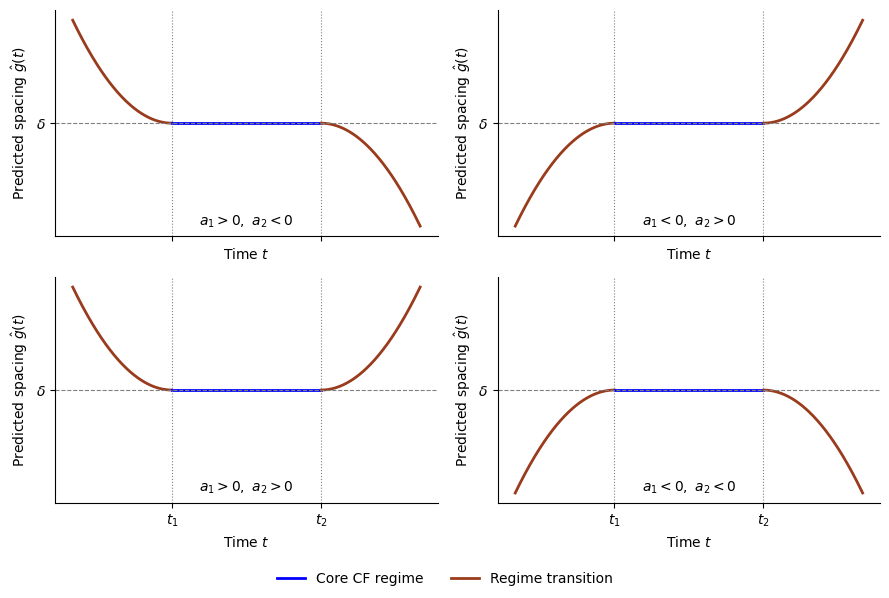}
    \caption{Possible configurations of predicted spacing in the extended model}
    \label{fig:ee model}
\end{figure}

To account for edge effects, the model is extended by introducing additional parameters. These parameters are calibrated by minimizing the RMSE between the observed and model-predicted spacings at the corresponding trajectory shift. In this formulation, each trajectory is partitioned into three segments: an initial regime transition phase, a core CF phase, and a final regime transition phase. Within the core CF segment, spacing is assumed to remain constant at $\delta$, consistent with the original model. In the transitional phases, spacing is allowed to vary over time to capture approach and divergence dynamics. The spacing is defined by piecewise quadratic structure-- a quadratic approach phase prior to time $t_1$, a steady plateau between $t_1$ and $t_2$, and a quadratic departure phase beyond $t_2$ (Eq. \ref{eq:piecewise model}). The steady region recovers the classical Newell spacing $\delta$. The observed shifted-spacing at shift $\tau$ for pair $k$ is defined as:
\begin{equation}
{g}_{\tau,k}(t) = x^{l}_{k}(t-\tau) - x^{f}_{k}(t)
\end{equation}
This is the quantity to which the model is fitted. The predicted shifted-spacing $\hat{g}_{\tau,k}(t)$ is given by:
\begin{equation}
\hat{g}_{\tau,k}(t) =
\begin{cases}
\delta + \dfrac{1}{2}\, a_1\,(t_1 - t)^2 & t < t_1 \quad \text{(entry regime transition)} \\[8pt]
\delta & t_1 \leq t \leq t_2 \quad \text{(core CF regime)} \\[8pt]
\delta + \dfrac{1}{2}\, a_2\,(t - t_2)^2 & t > t_2 \quad \text{(exit regime transition)}
\end{cases}
\label{eq:piecewise model}
\end{equation}
where $t_1$ and $t_2$ mark the boundaries of the core CF regime, and $a_1$ and $a_2$ are parameters governing the rate of change of spacing in the initial and final transitional phases, respectively. A positive value of $a_1$ implies that spacing increases as time moves away from $t_1$ toward $t = 0$, indicating that the follower is farther behind than the steady-state spacing during the entry phase. Similarly, a positive value of $a_2$ implies that spacing increases after $t_2$, indicating that the follower falls behind the leader during the exit phase (Fig. \ref{fig:ee model}). This formulation allows for a smooth transition between dynamic and steady-state behaviour, while preserving the interpretability of the core parameters. The extended formulation thus introduces additional degrees of freedom relative to the basic Newell formulation, increasing the total number of parameters to be estimated. In total, six parameters are estimated for each LF pair: the two original Newell parameters ($\tau$ and $\delta$ ), along with four additional parameters that characterize the edge effects.

\begin{equation}
\theta_k = (\tau, \delta, t_1, t_2, a_1, a_2),
\quad \text{subject to}
\quad
\begin{cases}
\tau < t_1 < t_2 < T \\
0 < \tau < 4 \\
1 < \delta < 15 \\
-5 < a_1 < 5 \\
-5 < a_2 < 5
\end{cases}
\end{equation}
$\theta_k$ is a parameter vector calibrated independently for each LF pair by minimizing the RMSE between the predicted $\hat{g}(t; \theta)$ and observed shifted spacing $g(t)$ series.

\noindent Objective function:
\begin{equation}
\hat{\theta}_k
=
\arg\min_{\theta}
\ \mathrm{RMSE}_k
=
\arg\min_{\theta}
\ \sqrt{
\frac{1}{N}
\sum_{i=1}^{N}
\left(g_{\tau,k}(t) - \hat{g}_{\tau,k}(t; \theta) \right)^2
}
\end{equation}
Calibration is performed using the Differential Evolution (DE) global optimisation algorithm \cite{storn1996usage}, which handles the non-convex objective surfaces arising from the piecewise structure and the discrete nature of the shift parameter $\tau$. The constraints are enforced as hard bounds during optimisation to ensure that a physically meaningful car-following interval exists within the observation window.

\begin{figure}[htbp]
\centering

\begin{subfigure}[b]{0.45\textwidth}
    \centering
    \includegraphics[width=\textwidth]{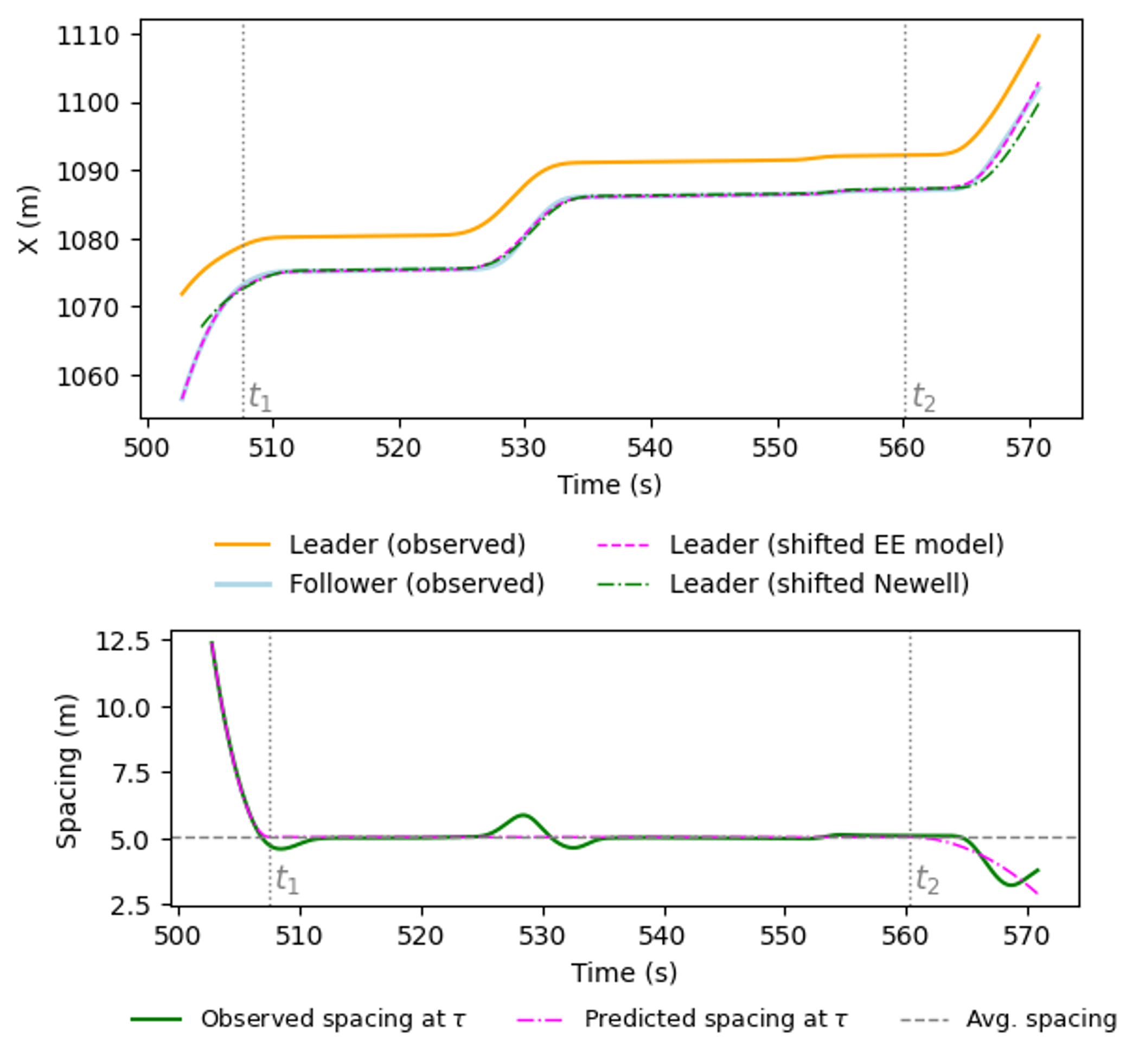}
    \caption{}
    \label{fig:ee ex 1}
\end{subfigure}
\hfill
\begin{subfigure}[b]{0.45\textwidth}
    \centering
    \includegraphics[width=\textwidth]{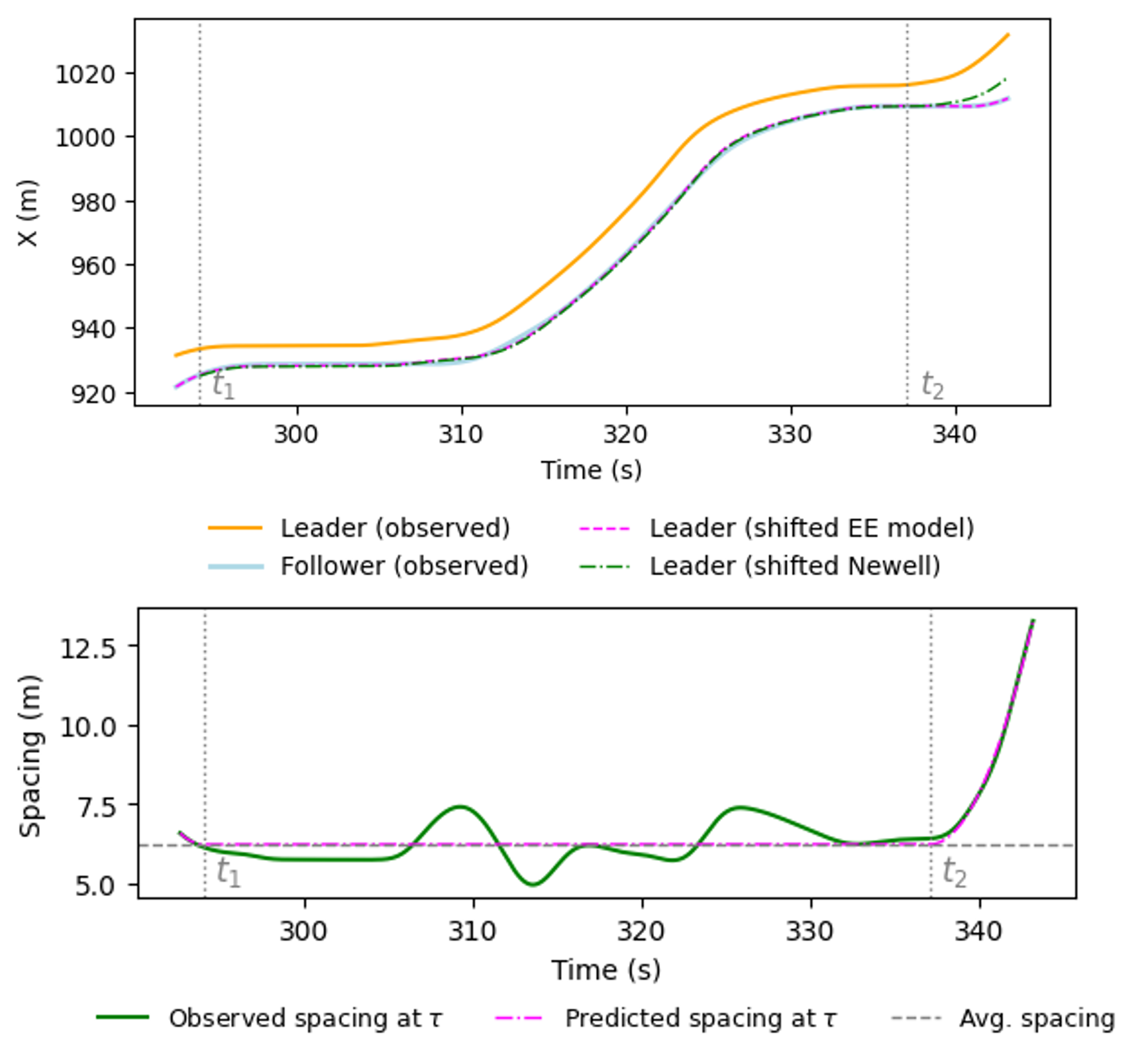}
    \caption{}
    \label{fig:ee ex 2}
\end{subfigure}
\caption{Trajectory and spacing fit comparison in the extended model}
\label{fig:ee ex main}
\end{figure}

The resulting six-parameter fits are compared against the basic two-parameter Newell model to assess whether the added parametric flexibility yields a statistically meaningful improvement in trajectory reproduction. The edge-effects model achieves a mean $\bar{R^2}$ of 0.95 and a mean RMSE of 0.4 m, substantially outperforming the basic shift. The gains are most pronounced in trajectory segments corresponding to the approach and divergence phases.
In the first example, edge effects appear at both ends of the trajectory. The simple Newell shift with $\tau=1.88s$ and $\delta=4.84m$ does not provide a satisfactory fit (Fig.~\ref{fig:ee1}), yielding an RMSE of 1.42 m and an $\bar{R}^2$ value of 0.55. When the extended formulation is applied, the fit improves markedly (Fig.~\ref{fig:ee ex 1}), yielding $\tau=1.16s$ and $\delta=5.04m$, with RMSE reduced to 0.23 m and $\bar{R}^2$ increased to 0.99. In the second example, where edge effects appear only near the latter portion of the trajectory (Fig.~\ref{fig:ee2}), the extended model again provides a superior fit (Fig.~\ref{fig:ee ex 2}); the RMSE decreases from 1.18 m to 0.54 m, while the $\bar{R}^2$ value increases from 0.85 to 0.97. The estimated parameters also adjust slightly, with $\tau$ changing from 1.44 s to 1.40 s and $\delta$ from 6.46 m to 6.24 m. Across all pairs, $\delta$ is largely stable between the two formulations, as it is anchored by the core CF segment. In contrast, the estimated $\tau$ value exhibits more pronounced variation across pairs, adjusting in conjunction with the additional parameters to minimize RMSE and optimize the overall trajectory fit.
These results attest to the effectiveness of the proposed extension, showing that relatively minor modifications to the original Newell model yield substantial gains in trajectory-fitting accuracy. By explicitly accounting for EE while retaining the core time-shift structure of Newell’s model, the proposed formulation achieves improved reconstruction fidelity without sacrificing conceptual parsimony.

\subsection{Model specifications}
The introduction of edge-effect terms increases the flexibility of the Newell framework, but also raises an important methodological question: whether the improvement in trajectory reproduction is sufficient to justify the additional model complexity. To examine this trade-off, a systematic model comparison was conducted using an information-theoretic framework that balances goodness-of-fit against model parsimony. Four candidate specifications of increasing structural complexity were evaluated for each LF pair (Table~\ref{tab:model_specs}).

\begin{table}[h]
\centering
\caption{Evaluated Model specifications}
\label{tab:model_specs}

\begin{tabular}{lllll}
\toprule
\textbf{Model} & \textbf{Description} & \textbf{Phases} & \textbf{Parameter vector ($\theta_k$)} & \textbf{$p$} \\
\midrule
S1 & Basic Newell & Constant only & $(\tau,\delta)$ & 2 \\
S2 & Leading-edge & Convergence → Steady & $(\tau,\delta,t_1,a_1)$ & 4 \\
S3 & Trailing-edge & Steady → Divergence & $(\tau,\delta,t_2,a_2)$ & 4 \\
S4 & Full edge-effect & Convergence → Steady → Divergence & $(\tau,\delta,t_1,t_2,a_1,a_2)$ & 6 \\
\bottomrule
\end{tabular}

\end{table}

\noindent The basic Newell model (S1) provides the most parsimonious representation, whereas the full edge-effect model (S4) offers the greatest flexibility to capture transitional spacing dynamics near the boundaries of the interaction interval. Although more flexible models generally reduce residual error, they are only preferable if the improvement in fit is large enough to compensate for the increased number of parameters.

Model parameters were estimated by minimizing the residual sum of squares (RSS) between observed and predicted spacing trajectories DE. The time offset parameter $\tau$ was determined through a discrete grid search over admissible temporal shifts, while the remaining parameters were optimized continuously. Accordingly, the effective parameter count $p$ used in model comparison included both the shift parameter $\tau$ and all continuously estimated model parameters. To compare competing specifications, the Bayesian Information Criterion (BIC) was computed for each fitted model using the least-squares forms derived under the assumption of independent Gaussian residual errors.

Lower values of BIC indicate a more favourable trade-off between model fit and model complexity. BIC imposes a stronger penalty on over-parameterized models, particularly for larger sample sizes, which is why it was adopted as the primary criterion for model selection. For each trajectory pair, the model with the lowest BIC value was identified as the preferred specification.

The comparison results show that the full edge-effect specification (S4) was selected for 58 of the 72 analyzed LF pairs, suggesting that the majority of pairs warrant boundary correction at both ends of the observation window. In these cases, the reduction in residual error achieved by incorporating both leading- and trailing-edge effects was sufficient to outweigh the additional complexity penalty imposed by BIC. This indicates that EE at both the initiation and termination of car-following interactions contribute materially to the observed trajectories.
Of the remaining 14 pairs, 12 were best represented by S2 specification and two by S3 specification. For these pairs, the added flexibility of the S4 did not yield sufficient improvement in fit to justify the extra parameters, suggesting that edge-effects were concentrated at a single boundary of the observation interval. Notably, of the 12 pairs for which S2 was selected, eight corresponded to trajectories that terminated naturally at the trailing edge of the observation window. This means the interaction did not end at that boundary, so no trailing-edge effect was present to correct for. This pattern lends further support to the interpretation that model selection is tracking genuine physical boundary effects rather than artifacts of overfitting.

These results indicate that deviations from the classical Newell formulation are not purely random residual fluctuations, but are systematically associated with transient phases of LF interactions. At the same time, it is important to note that the superiority of the extended model does not invalidate the core assumptions of the Newell framework. Rather, it highlights the conditions under which these assumptions are most applicable. The basic model remains an appropriate representation of steady-state following behaviour, while the extended formulation provides a more comprehensive description that includes transitional dynamics.

The model comparison results suggest that a hybrid perspective is necessary for accurately representing CF behaviour in mixed traffic conditions. A purely constant-parameter model is insufficient to capture the full range of observed dynamics, but a modest extension that accounts for edge effects can significantly improve model performance without sacrificing interpretability. This provides a practical pathway for enhancing the applicability of Newell-type models in complex traffic environments.

\section{Results} \label{sec:8}
\subsection{Reduction of Boundary-Induced Variance}
In the speed–spacing plots, the points exhibiting the largest deviations from the regression line, and thus contributing disproportionately to the observed variance, are systematically eliminated once edge effects are accounted for. These outlying observations correspond to segments lying outside the core CF regime. Overall, approximately 55\% of the datapoints fall within the core CF regime, with the remainder attributable to regime-transition behaviour. When the regression is restricted to core CF points alone, $R^2$ improves to 0.60, a modest gain over the 0.54 obtained for the aggregate dataset. This improvement suggests that some share of the observed scatter arises not only from behavioural heterogeneity, but also from the inclusion of regime-transition segments. Correcting for EE thus yields a meaningful reduction in unexplained variance, indicating that this step should be treated as an integral component of any LF analysis.

This finding has further implications beyond variance reduction. LF pair identification is conventionally performed using heuristic thresholds (e.g., fixed spacing or time-gap cutoffs), an approach with well-documented limitations \cite{kulkarni2025leader}. By explicitly recovering the boundaries of the core CF regime, edge-effect correction offers a principled, data-driven alternative to such heuristics: rather than imposing thresholds \textit{a priori}, the actual extent of sustained car-following engagement can be inferred directly from the trajectory data, enabling more reliable extraction of genuine LF pairs.

\subsection{Model Comparisons}
The trajectory-shifting method and regression-based models represent two methodologically distinct approaches to evaluating the validity of Newell’s CF framework. Whereas the regression-based approach assesses model validity indirectly through the statistical relationship between speed and spacing, the trajectory-shifting method provides a more direct test of Newell’s constant-shift hypothesis.
A systematic comparison of alternative specifications that are conceptually and empirically comparable, both within and across these two approaches, is presented here.
This comparison enables rigorous cross-validation of parameter estimates and offers a more comprehensive basis for evaluating the degree to which the behavioural assumptions underlying the model are supported by empirical observations.

\subsubsection{Simple Trajectory-Shifting vs.\ Simple Regression}
This subsection compares Newell's parameters estimated by pair-wise regression with leader speed as predictor (LR) and simple trajectory-shifting (TS). The TS method achieves moderate agreement between predicted and observed trajectories, with a mean variance-reduction index $\bar{R^2}=0.67$, indicating that the time-shifted leader trajectory accounts for a substantial portion of the spacing dynamics in a typical pair. This performance compares favourably with the LR model, yielding a mean $R^2$ of 0.63. It is important to interpret these figures correctly: both metrics are arithmetic means of pair-level $R^2$ values and therefore characterize goodness-of-fit (GOF) for a representative pair. They are conceptually distinct from the pooled $R^2$ reported earlier, which quantifies the proportion of total variance explained across all observations jointly, referenced against the grand mean spacing rather than individual pair means.

\begin{figure}[H]
\centering

\begin{subfigure}[b]{0.45\textwidth}
    \centering
    \includegraphics[width=\textwidth]{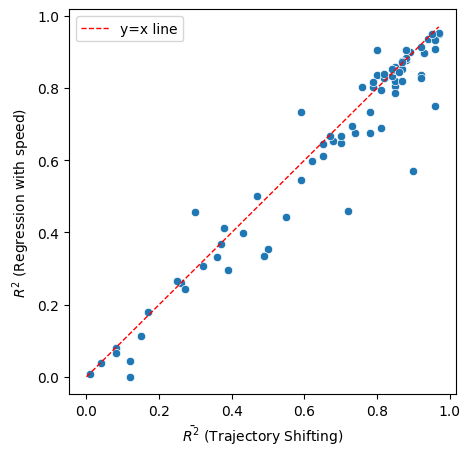}
    \caption{}
    \label{fig:R21}
\end{subfigure}
\hfill
\begin{subfigure}[b]{0.45\textwidth}
    \centering
    \includegraphics[width=\textwidth]{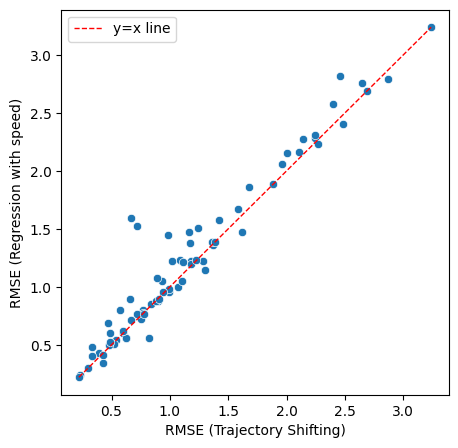}
    \caption{}
    \label{fig:rmse1}
\end{subfigure}
\caption{GOF comparison of simple trajectory shifting and simple regression approach}
\label{fig: comparison11}
\end{figure}

The scatter plots in Fig.~\ref{fig: comparison11} collectively provide cross-methodological evidence for the validity of Newell's simplified CF model. Figure~\ref{fig:R21} reveals a strong positive association between the $R^2$ values obtained from the TS and LR approach, with the majority of pair-wise observations clustering close to the 1:1 reference line. A discernible tendency for points to fall below this line indicates that the TS approach yields marginally superior explanatory power for most of the pairs. This finding is also supported by the higher mean $\bar{R^2}$ value obtained under the TS approach and is likely attributable to its more direct formulation. Figure~\ref{fig:rmse1} corroborates this pattern through the lens of RMSE, where the near-diagonal scatter of points indicates that the spacing prediction errors generated by both approaches are broadly comparable across the full sample. Taken together, these convergent GOF profiles, obtained through two methodologically independent routes, lend substantial support to the proposition that the linear spacing–speed relationship posited by Newell's model adequately characterizes CF behaviour in the study context. The consistency observed across pairs spanning a wide performance range further suggests that the model's validity is not confined to well-behaved, high-$R^2$ pairs, but extends, at least approximately, to the more heterogeneous follower behaviour documented in the lower-fit observations.

\begin{figure}[H]
\centering

\begin{subfigure}[b]{0.45\textwidth}
    \centering
    \includegraphics[width=\textwidth]{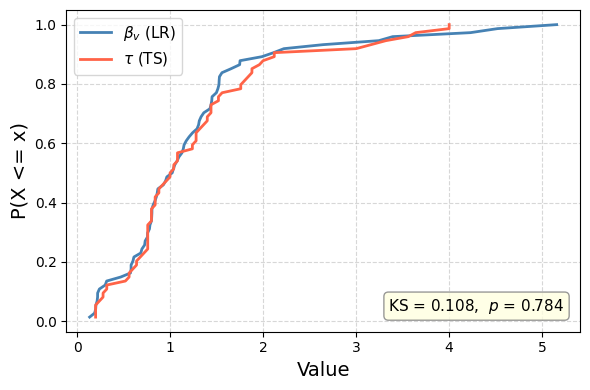}
    \caption{}
    \label{fig:tau1}
\end{subfigure}
\hfill
\begin{subfigure}[b]{0.45\textwidth}
    \centering
    \includegraphics[width=\textwidth]{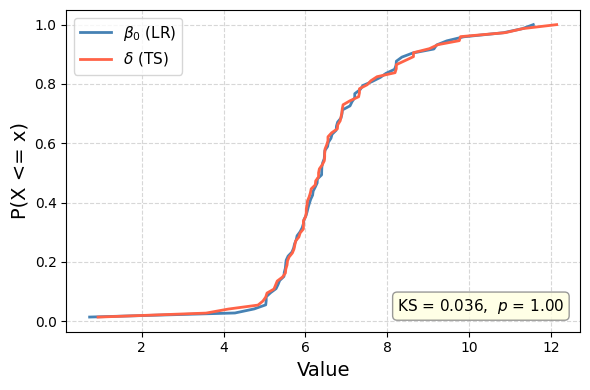}
    \caption{}
    \label{fig:delta1}
\end{subfigure}
\caption{Distribution comparison of $\tau$ and $\delta$ from simple LR and trajectory shifting}
\label{fig: comparison12}
\end{figure}
The empirical CDFs presented in Fig.~\ref{fig: comparison12} provide a distributional comparison of the Newell model parameters estimated through two independent approaches. For the time offset parameter (Fig.~\ref{fig:tau1}), a somewhat larger but still modest KS statistic of 0.11 (p = 0.77) is observed between the regression speed coefficient $\beta_v$ and the time-offset estimate $\tau$. The null hypothesis of distributional equivalence cannot be rejected at any conventional significance level, and the two ECDFs converge substantially up to the second quartile with a median of approximately 1~s. For the jam spacing parameter (Fig.~\ref{fig:delta1}), the distributions of the regression intercept $\beta_0$ and the shifted spacing $\delta$ are, to a remarkable degree, indistinguishable: the two ECDFs trace virtually the same path across the full support of the data. This is also confirmed by a K–S statistic of 0.04 and an associated p-value of $\approx1.0$, indicating no statistically significant difference between the two distributions. The distributions thus have coincident medians of approximately 6.2~m. This near-perfect agreement lends strong support to interpreting both $\beta_0$ and $\delta$ as consistent empirical surrogates for the jam-spacing construct embedded in Newell's model. Collectively, these results demonstrate that the two estimation frameworks recover statistically equivalent parameter populations, reinforcing the structural consistency of Newell's linear CF relationship and validating the trajectory-shifting approach as a simple and intuitive alternative to regression-based calibration.

\subsubsection{Trajectory-Shifting with and without Edge-Effect Correction}
This subsection compares the parameters of the Newell model estimated using the simple trajectory-shifting (TS) and the trajectory-shifting with edge-effects (TSEE) approaches. Accounting for edge effects leads to a substantial improvement in model performance. The mean variation-reduction index ($\bar{R^2}$) increases from 0.67 for TS to 0.95 for TSEE, indicating a markedly better fit to the observed trajectories. Similarly, the RMSE decreases from 1.2 m to 0.40 m, demonstrating a significant reduction in prediction error.
\begin{figure}
\centering

\begin{subfigure}[b]{0.45\textwidth}
    \centering
    \includegraphics[width=\textwidth]{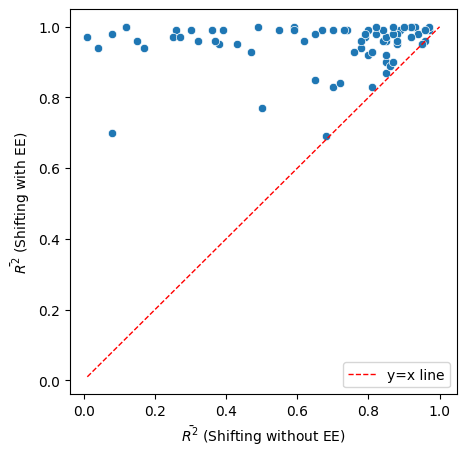}
    \caption{}
    \label{fig:R22}
\end{subfigure}
\hfill
\begin{subfigure}[b]{0.45\textwidth}
    \centering
    \includegraphics[width=\textwidth]{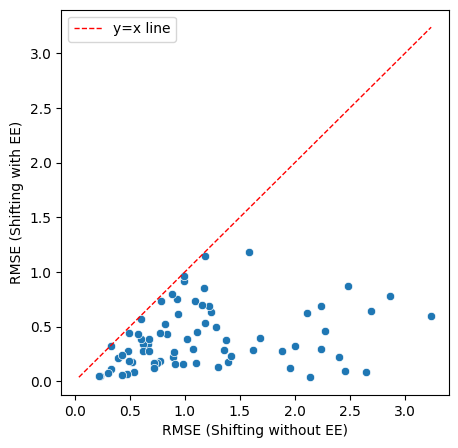}
    \caption{}
    \label{fig:rmse2}
\end{subfigure}
\caption{Effect of EE Correction on GOF}
\label{fig: comparison21}
\end{figure}
The scatter plots in Fig.~\ref{fig: comparison21} reveal a pronounced and systematic improvement in model fit when the EE correction is applied to the trajectory-shifting procedure, with the direction and magnitude of change being strikingly consistent across the entire sample. In  Fig.~\ref{fig:R22}, the overwhelming majority of points lie well above the 1:1 reference line, indicating that EE-corrected $\bar{R^2}$ values are substantially higher than their uncorrected counterparts; notably, a large cluster of pairs attains $\bar{R^2}$ values approaching unity under the corrected scheme even when the uncorrected $\bar{R^2}$ is near zero, underscoring that the poor fit observed without correction is largely an artefact of edge-effects rather than a genuine failure of the model to describe the underlying CF dynamics.  Fig. \ref{fig:rmse2} reinforces this interpretation through RMSE. Almost all observations fall markedly below the reference line, with pairs that exhibited RMSE values in the range of 1.5–3.2 m under the uncorrected approach being reduced to well below 1.0 m once the correction is applied. This significant reduction is consistent with the expected behaviour due to the edge-effects. The near-universal nature of the improvement, with only negligible exceptions, suggests that "edge-effects" is a pervasive feature of finite CF interactions in this dataset, and that its correction is not merely beneficial but essential for an unbiased assessment of Newell model validity under trajectory-shifting.
\begin{figure}[H]
\centering

\begin{subfigure}[b]{0.45\textwidth}
    \centering
    \includegraphics[width=\textwidth]{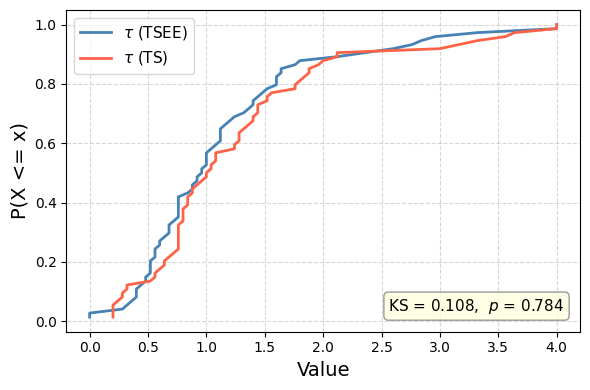}
    \caption{}
    \label{fig:tau2}
\end{subfigure}
\hfill
\begin{subfigure}[b]{0.45\textwidth}
    \centering
    \includegraphics[width=\textwidth]{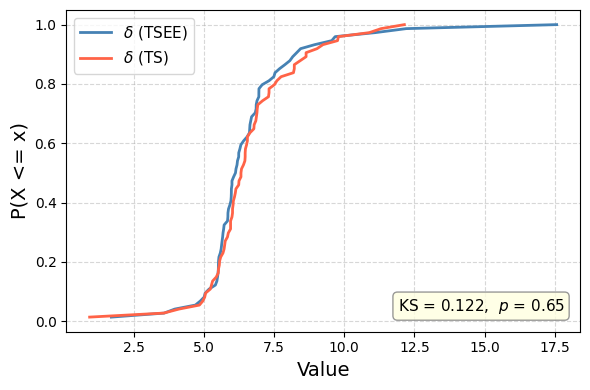}
    \caption{}
    \label{fig:delta2}
\end{subfigure}
\caption{Effect of EE Correction on Parameter Distributions}
\label{fig: comparison22}
\end{figure}
The ECDFs in Fig.~\ref{fig: comparison22} illustrate how the boundary correction influences the distributional characteristics of the estimated model parameters, offering insight into the systematic bias introduced by boundary contamination in the TS approach. For the $\tau$ parameter (Fig.~\ref{fig:tau2}), a more substantively interpretable shift is observed. The TSEE distribution accumulates faster in the lower range (approximately 0.3–2 s), indicating that EE correction systematically reduces time lag estimates for a notable fraction of pairs, with the TSEE curve lying consistently to the left of the TS curve almost for the entire range. This reduction is mechanically plausible, as unaccounted edge-effects can artificially inflate spacing measurements near the boundaries of the observation window. Consequently, the variance-minimizing trajectory shift may be biased toward larger values of $\tau$, leading to an overestimated value. Despite this directional shift, the KS statistic of 0.08 (p = 0.97) indicates that the two $\tau$ distributions remain statistically indistinguishable.  This suggests that while EE correction improves parameter estimation at the individual-pair level, it does not substantially alter the overall representation of pair time offset heterogeneity within the sample. For the jam spacing parameter $\delta$ (Fig.~\ref{fig:delta2}), the TSEE distribution exhibits a slight leftward shift relative to the TS distribution. However, the corresponding K--S statistic of 0.14 ($p = 0.49$) confirms that this difference is not statistically significant, indicating that EE correction has only a modest effect on the aggregate distribution of estimated jam spacing values.

\subsubsection{Simple vs.\ Multiple Linear Regression}
This subsection compares the parameters estimated by simple linear regression with speed as the sole explanatory variable (LR), and multiple linear regression (MLR), which includes speed, acceleration, and deceleration. The inclusion of acceleration-related terms results in a noticeable improvement in model performance. The mean pairwise $R^2$ increases from 0.63 for LR to 0.75 for MLR, indicating a substantially better representation of the spacing dynamics. Correspondingly, the RMSE decreases from 1.26~m to 1.04~m, reflecting a significant reduction in prediction error.
\begin{figure}[H]
\centering

\begin{subfigure}[b]{0.45\textwidth}
    \centering
    \includegraphics[width=\textwidth]{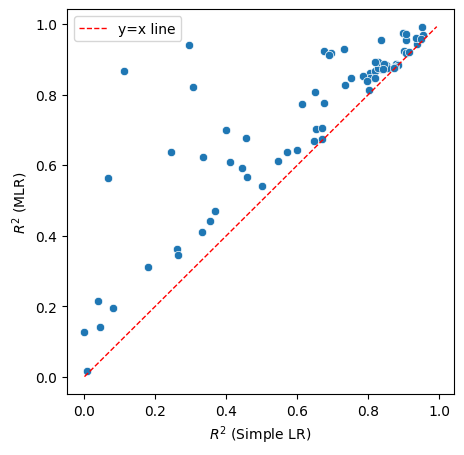}
    \caption{}
    \label{fig:R23}
\end{subfigure}
\hfill
\begin{subfigure}[b]{0.45\textwidth}
    \centering
    \includegraphics[width=\textwidth]{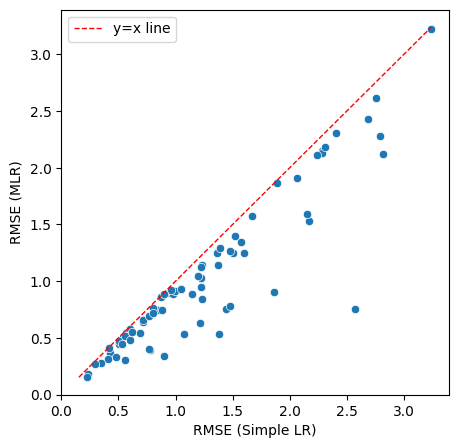}
    \caption{}
    \label{fig:rmse3}
\end{subfigure}
\caption{Simple Linear Regression vs. Multiple Linear Regression: GOF Comparison}
\label{fig: comparison31}
\end{figure}
The scatter plots in Fig.~\ref{fig: comparison31} document a systematic and widespread improvement in GOF when the pairwise MLR specification is employed in place of the simple LR, with the directional advantage of MLR manifest across the full range of observed data. In Fig.~\ref{fig:R23}, the near-universal disposition of points above the 1:1 reference line indicates that MLR consistently recovers higher $R^2$ values, with the improvement being most pronounced for pairs that exhibit poor fit under LR-- pairs with $R^2$ values close to zero under the simpler specification attain values in the range of 0.15–0.85 under MLR, indicating that the additional regressors capture systematic variance in spacing behaviour that the single-predictor model cannot accommodate. Fig.~\ref{fig:rmse3} corroborates this pattern by showing that the majority of points fall below the reference line, confirming that MLR yields lower prediction errors across the sample. The improvement seems broadly proportional in magnitude, with the largest absolute reductions occurring among pairs with the highest RMSE under LR. These results suggest that the linear Newell spacing–speed relationship does not fully exhaust the explainable variance in observed spacing for a non-trivial subset of pairs under its simplest formulation. The MLR framework, by accommodating additional sources of systematic variation (whether through pooled fixed effects, pair-level heterogeneity in intercepts and slopes, or supplementary predictors), provides a more accurate representation of the CF dynamics. This improvement, however, must be interpreted alongside the significant increase in model complexity, and the trade-off between parsimony and explanatory adequacy warrants consideration when selecting a specification for predictive applications.

\begin{figure}[H]
\centering

\begin{subfigure}[b]{0.45\textwidth}
    \centering
    \includegraphics[width=\textwidth]{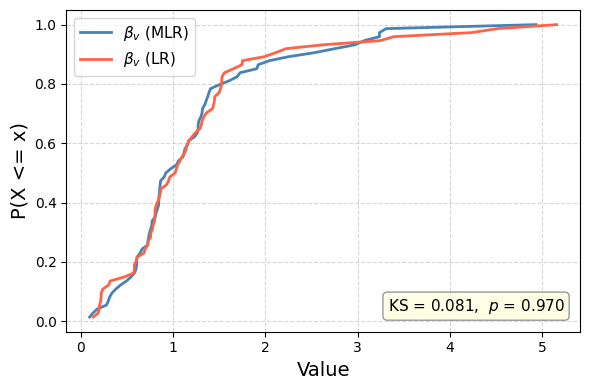}
    \caption{}
    \label{fig:tau3}
\end{subfigure}
\hfill
\begin{subfigure}[b]{0.45\textwidth}
    \centering
    \includegraphics[width=\textwidth]{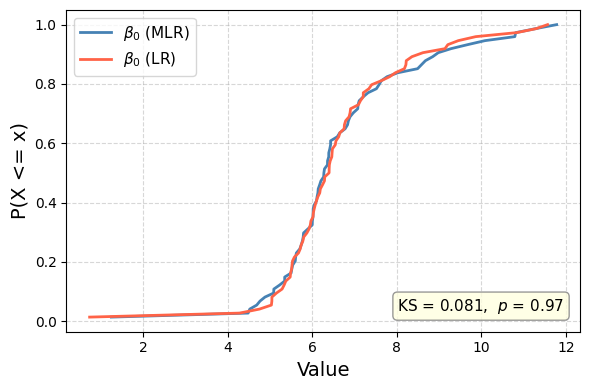}
    \caption{}
    \label{fig:delta3}
\end{subfigure}
\caption{Distribution comparison of $\tau$ and $\delta$ from simple and multiple linear regression}
\label{fig: comparison32}
\end{figure}
It must be noted here that $\beta_v$ and $\beta_0$ from regression are taken as estimates of $\tau$ and $\delta$, respectively. A particularly noteworthy finding emerges from the comparison of parameter distributions between the two specifications. Despite the demonstrably superior GOF of the MLR approach, as documented in  Fig.~\ref{fig: comparison31}, the two frameworks recover statistically equivalent parameter populations for both Newell parameters. 
The almost identical distributions of $\beta_v$ (Fig.~\ref{fig:tau3}) in both models, exhibiting only minor local deviations in the upper quartile region, show that the speed coefficient remains stable even after adding acceleration and deceleration variables. This supports the theoretical basis of Newell’s CF model, where the spacing-speed relationship is linear, and its slope represents the time offset $\tau$. An equally strong result is obtained for the jam spacing intercept $\beta_0$ (Fig.~\ref{fig:delta3}), where the LR and MLR ECDFs are virtually superimposed across the entire value range. The identical KS statistics confirm that the observed divergence is well within the range of sampling variability, providing no grounds whatsoever to reject distributional equivalence. The juxtaposition of these distributional results with the GOF comparisons carries a substantively important implication. The improved explanatory power of MLR arises from its capacity to account for systematic between-pair heterogeneity and additional sources of spacing variance. Notably, this enhancement is achieved without introducing substantial changes in the estimates of the speed coefficient ($\beta_v$) and intercept ($\beta_0$), increasing confidence in their interpretation as reliable estimators of the parameters $\tau$ and $\delta$, respectively. In other words, the structural parameters of the model are robustly identified regardless of regression specification, lending confidence that the parameter distributions reported in this study reflect genuine behavioural characteristics of the observed CF
pairs rather than artefacts of the chosen estimation framework.

Across all three comparisons, the jam-spacing parameter $\delta$, whether estimated as $\beta_o$ (LR or MLR) or $\delta$ (shifting, with or without EE correction), exhibits consistent near-distributional identity, with KS statistics uniformly below 0.12 and p-values uniformly above 0.5. This cross-method concordance suggests that all four estimation frameworks converge on the same underlying physical quantity, i.e., the jam spacing, lending credibility to the overall parameter identification strategy. The robustness of $\delta$ to model specification provides support for jam spacing as a stable anchor in the Newell parameter space and verifies the internal consistency of the estimation pipeline developed in this study.

\section{Discussion} \label{sec:9}
The results provide a nuanced assessment of Newell's CF model under lane-free traffic conditions.  Rather than a binary verdict of validity or failure, the evidence suggests that the applicability of the model is conditional on two factors: the treatment of driver heterogeneity and the presence of edge-effects due to regime transitions. Thus, the regression-based and trajectory-based analyses indicate that the core behavioural principles underlying Newell's formulation remain observable in the data, but their explanatory power improves substantially when these factors are explicitly accounted for.

\subsection{Driver Heterogeneity and Validity of Speed–Spacing Relationship}
One of the clearest findings of this study is the dominant role of inter-driver heterogeneity in determining spacing behaviour. The progression from aggregate models to pair-specific specifications produced a substantial increase in explanatory power, with the pooled coefficient of determination increasing from approximately 0.51 under the aggregate linear model to 0.84 under the pair-specific model incorporating asymmetric acceleration effects. This improvement suggests that a large portion of the variance unexplained by aggregate formulations arises from systematic behavioural differences among drivers rather than random fluctuations. From the perspective of Newell's theory, this finding is important because the model assumes that each driver is characterized by a unique pair of parameters ($\tau, \delta$). The present results provide empirical support for this interpretation under conditions substantially more heterogeneous than those originally considered in the development of the model. The findings suggest that $\tau$ and $\delta$ are better viewed as behavioural characteristics that vary across drivers rather than as universal constants applicable to an entire traffic stream, indicating that behavioural heterogeneity remains a fundamental characteristic even in highly disordered traffic environments.

The comparatively modest improvement achieved by introducing acceleration and deceleration variables after accounting for pair-specific effects also offers behavioural insight. While acceleration asymmetry contributes additional explanatory power, speed remains the dominant determinant of spacing, indicating that the linear speed–spacing relationship proposed by Newell continues to capture the primary structure of CF behaviour at the level of individual pairs. However, this limited marginal contribution to spacing prediction should not be read as evidence that acceleration and deceleration effects are behaviourally inconsequential. Asymmetries of this kind, even where small in magnitude, are known to aggregate into macroscopic phenomena such as hysteresis in the speed–spacing plane and stop-and-go wave propagation, effects that are not necessarily visible at the level of this analysis. Disentangling these scale-dependent implications, microscopic versus macroscopic, falls outside the scope of the present analysis and is identified as a direction for future work.

\subsection{Edge Effects and the Behavioural Scope of Newell's Model}
The most consequential finding of the study concerns the role of edge-effects. The substantial improvement in trajectory reproduction achieved after introducing boundary corrections, together with the overwhelming preference for the full edge-effect specification under the BIC criterion, indicates that deviations from the classical Newell model are concentrated primarily at the initiation and termination of car-following interactions. This observation has important theoretical implications. The results suggest that many apparent failures of the Newell framework arise not because the underlying trajectory-translation principle is fundamentally invalid, but because real-world interactions contain transitional regimes that fall outside the behavioural scope of the original model. During approach and divergence phases, drivers actively adjust spacing to establish or dissolve a following relationship, violating the steady-state assumptions implicit in a constant pair of parameters. When these transitional phases are explicitly represented, the explanatory performance of the trajectory-based framework improves dramatically.

From this perspective, the edge-effect formulation should not be viewed as a replacement for Newell's model but rather as a mechanism for delimiting its behavioural domain. The central portion of most interactions is well described by the classical time–space translation framework, whereas the approach and disengagement phases call for additional modelling flexibility. Prior studies have generally addressed such discrepancies by relaxing the steady-state assumptions underlying Newell-type models, typically through stochastic or dynamic extensions. The present findings, however, suggest that a substantial portion of the observed discrepancy may originate from the finite interaction boundaries themselves, offering an alternative explanation for deviations often attributed to stochastic driver behaviour.

\subsection{Implications for Modelling Mixed Traffic}
The findings support a conditional interpretation of Newell's CF model in the observed traffic environment. The evidence suggests that the model remains capable of reproducing key aspects of observed CF behaviour, particularly during sustained following periods, provided that driver-specific parameter variation is acknowledged and transitional boundary effects are treated separately. Under these conditions, both the speed–spacing relationship and the trajectory-translation principle receive substantial empirical support. The results, therefore, point towards a hybrid modelling perspective. A strictly constant-parameter representation is insufficient to capture the full range of observed behaviour, yet the core structure of Newell's model remains valuable because of its behavioural interpretability and parsimony. Modest extensions that account for driver heterogeneity and interaction-boundary dynamics appear capable of substantially improving performance while preserving the conceptual simplicity that has made the Newell framework attractive for traffic-flow applications. Such an approach may provide a practical pathway for extending Newell-type models to complex mixed-traffic environments without abandoning their underlying theoretical foundation.

\section{Limitations and Future Directions} \label{sec:10}
The proposed framework advances the evaluation of Newell's CF model under heterogeneous, lane-free traffic conditions, yet several limitations constrain its current scope and point to productive directions for future work.

\begin{figure}[htbp]
\centering
\begin{subfigure}[b]{0.45\textwidth}
    \centering
    \includegraphics[width=\textwidth]{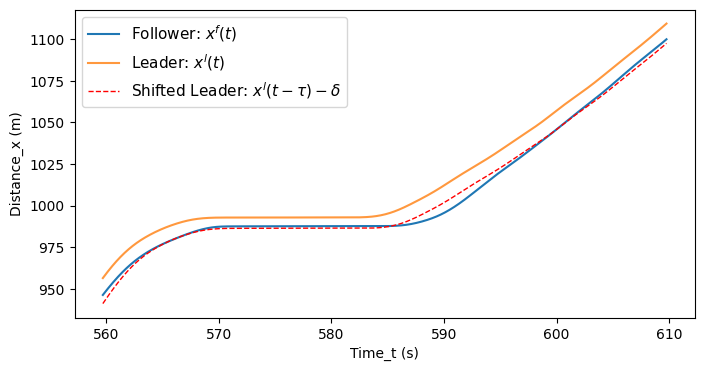}
    \label{fig:ade1}
\end{subfigure}
\hfill
\begin{subfigure}[b]{0.45\textwidth}
    \centering
    \includegraphics[width=\textwidth]{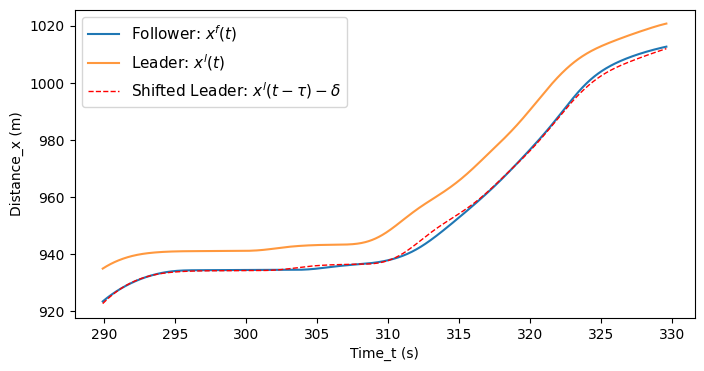}
    \label{fig:ade2}
\end{subfigure}
\caption{Acceleration/ deceleration effects}
\label{fig:ade}
\end{figure}

\begin{figure} [htbp]
    \centering
    \includegraphics[width=0.5\linewidth]{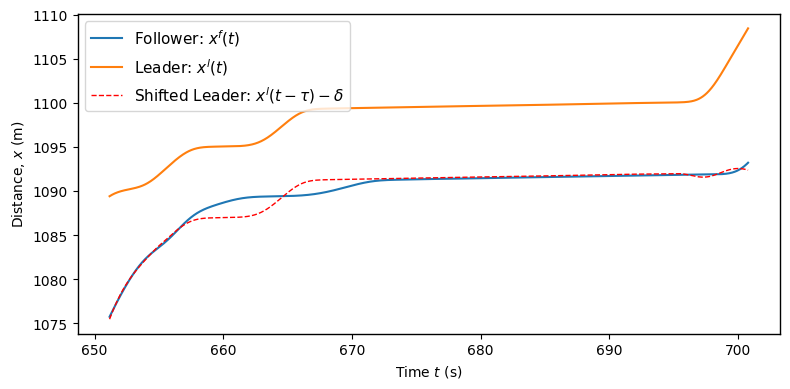}
    \caption{Irregular behaviour}
    \label{fig:car100}
\end{figure}
\begin{enumerate} [label=\roman*)]
    \item  The edge-effect correction addresses the initiation and termination phases of leader-follower interactions, but the intermediate following regime is still governed by constant temporal and spatial translations. In practice, sudden acceleration and deceleration generate residual discrepancies that a constant-translation model cannot resolve, as Fig.~\ref{fig:ade} illustrates. Occasional irregular trajectories, as in Fig.~\ref{fig:car100}, fall outside the model's scope entirely and most likely reflect driver responses to conditions further downstream rather than to the immediate leader. Whether these types of cases can be incorporated into the trajectory-shifting framework without undermining its interpretability is an open question.

    \item The framework also treats each interaction as governed by a single, time-invariant pair of Newell parameters throughout the steady-state window. This assumption may hold for well-defined following episodes, but in practice, drivers adjust their spacing and response timing as traffic conditions and driving intent change. Adaptive or state-dependent estimation approaches that allow parameters to evolve over the course of an interaction, while remaining behaviourally interpretable, are a natural extension.

    \item A more immediate methodological concern involves the LF identification criteria. The study classifies a vehicle as a follower whenever any lateral overlap with the leader exists, however small. In lane-free conditions, the degree of overlap may matter: a vehicle with only marginal overlap occupies a materially different position from one tracking closely behind its leader, and the two situations may produce different spacing and headway characteristics. Additionally, a vehicle can have a partial lateral overlap with more than one vehicle. How sensitive the estimated Newell parameters are to the extent of overlap or multiple leaders has not been examined here.

    \item The analysis is further restricted to car pairs. Mixed traffic includes two-wheelers, auto-rickshaws, and heavy vehicles alongside passenger cars, each with different dimensions, performance capabilities, and gap-acceptance behaviour. Whether Newell-type models can describe these cross-vehicle interactions, and whether the edge-effect correction generalizes to them, remain open questions. More fundamentally, Newell's model is a single-leader formulation. In dense mixed traffic, the following behaviour is frequently shaped by multiple vehicles ahead and by relative lateral position in the stream. Extending the trajectory-shifting framework to multi-leader or two-dimensional settings would address this limitation, though it would require substantially different parameterization.

    \item The regression analysis does not account for residual autocorrelation, which may inflate the reported $R^2$ values. Addressing autocorrelation should be considered in future work.
    
    \item The trajectory length in this study is limited by the approximately 180 m road segment covered by a single drone, resulting in a limited number of CF interactions. Future work will extend the analysis to longer trajectories by stitching together data from multiple drones.

    \item Finally, all data come from a single study site. The particular geometry, composition, and congestion level of that location may affect both the parameter distributions and the degree of edge-effect distortion observed. Replication with other well-known datasets, such as NGSIM, is needed before the findings can be considered generalizable.
\end{enumerate}
   
\noindent The core finding remains that, after correcting for edge effects, Newell's model provides a tractable representation of longitudinal car-following behaviour even in heterogeneous, lane-free traffic. The stated constraints delimit the scope of this study, not the inherent applicability of the model.

\section*{Data Availability}
\label{sec:data}
The trajectory dataset developed in this study is available for 
non-commercial research purposes upon reasonable request.

\section*{CRediT Authorship Contribution Statement}
\textbf{Suhaib Nazir:} Conceptualization, Methodology, Formal 
analysis, Writing -- original draft. 
\textbf{Hillel Bar-Gera:} Conceptualization, Supervision, Writing 
-- review \& editing. 
\textbf{Bhargava Rama Chilukuri:} Supervision, Funding acquisition, 
Writing -- review \& editing.

\section*{Acknowledgments}
The authors thank project associates K.~Vijayraj, E.~Raji, Bhuvana, 
and Vishnupriya for their assistance with data collection and 
extraction.

\bibliographystyle{splncs04}
\bibliography{references}

\end{document}